\pdfoutput=1
\documentclass[cernpreprint, PAPER, atlasdraft=false, texlive=2023, UKenglish, texmf, orcidlogo]{atlasdoc}
 
\usepackage{atlaspackage}
\usepackage{atlasbiblatex}
\usepackage{cancel}
\usepackage{verbatim}
\usepackage{multirow}
\usepackage{makecell}
 
\usepackage{atlasphysics}
\usepackage{atlascover}
\usepackage{float}
\usepackage{atlasphysics}

\graphicspath{{logos/}{figures/}}
 
\usepackage{ANA-HDBS-2022-44-PAPER-defs}

\AtlasTitle{Search for a new heavy scalar particle decaying into a Higgs boson and a new scalar singlet in final states with one or two light leptons and a pair of $\tau$-leptons with the ATLAS detector}

\AtlasAbstract{
A search for a new heavy scalar particle $X$ decaying into a Standard Model (SM) Higgs boson and a new singlet scalar particle $S$ is presented. The search uses a proton--proton ($pp$) collision data
sample with an integrated luminosity of 140 \ifb\,recorded at a centre-of-mass energy of $\sqs = 13$ \TeV\,with the ATLAS detector at the Large Hadron Collider. The most sensitive mass parameter space
is explored in $X$ mass ranging from 500 to 1500 \GeV, with the corresponding $S$ mass in the range 200--500 \GeV.
The search selects events with two  hadronically decaying $\tau$-lepton candidates from $H\to \tau^+\tau^-$ decays and one or two light leptons ($\ell=e,\,\mu$) from $S\to VV$
($V = W,\,Z$) decays while the remaining $V$ boson decays hadronically or to neutrinos. A multivariate discriminant based on event kinematics is used to  separate the signal from the background.
No excess is
observed beyond the expected SM background and 95\% confidence level
upper limits between 72~fb and 542~fb are derived on the cross-section $\sigma(pp\to X\to SH)$ assuming
the same SM-Higgs boson-like decay branching ratios for the $S\to VV$ decay.
Upper limits on the visible cross-sections $\sigma(pp\to X\to SH \to WW\tau\tau)$
and $\sigma(pp\to X\to SH \to ZZ\tau\tau)$ are also set in the ranges 3--26~fb and 6--33~fb, respectively.
 
}

\AtlasRefCode{HDBS-2022-44}
 
\PreprintIdNumber{CERN-EP-2023-125}

\AtlasJournalRef{JHEP 10(2023) 009}
\AtlasDOI{10.1007/JHEP10(2023)009}

\AtlasJournal{JHEP}

\hypersetup{pdftitle={ATLAS document},pdfauthor={The ATLAS Collaboration}}
 
\begin{document}
 
\maketitle
 
\tableofcontents

\clearpage

\section{Introduction}
\label{sec:intro}
 
The discovery of the Higgs boson ($H$) in 2012~\cite{HIGG-2012-27, CMS-HIG-12-028} by the ATLAS~\cite{PERF-2007-01} and
CMS~\cite{CMS-CMS-00-001} collaborations at the Large Hadron Collider (LHC) has led to a
comprehensive programme of measurements and searches using proton--proton ($pp$) collision data.
All measurements to date are consistent with the prediction of the Standard Model (SM)~\cite{HIGG-2015-07, HIGG-2021-23, CMS-HIG-22-001,CMS:2022suh}.
 
\begin{figure}[htbp]
\centering
\includegraphics[width=0.6\textwidth]{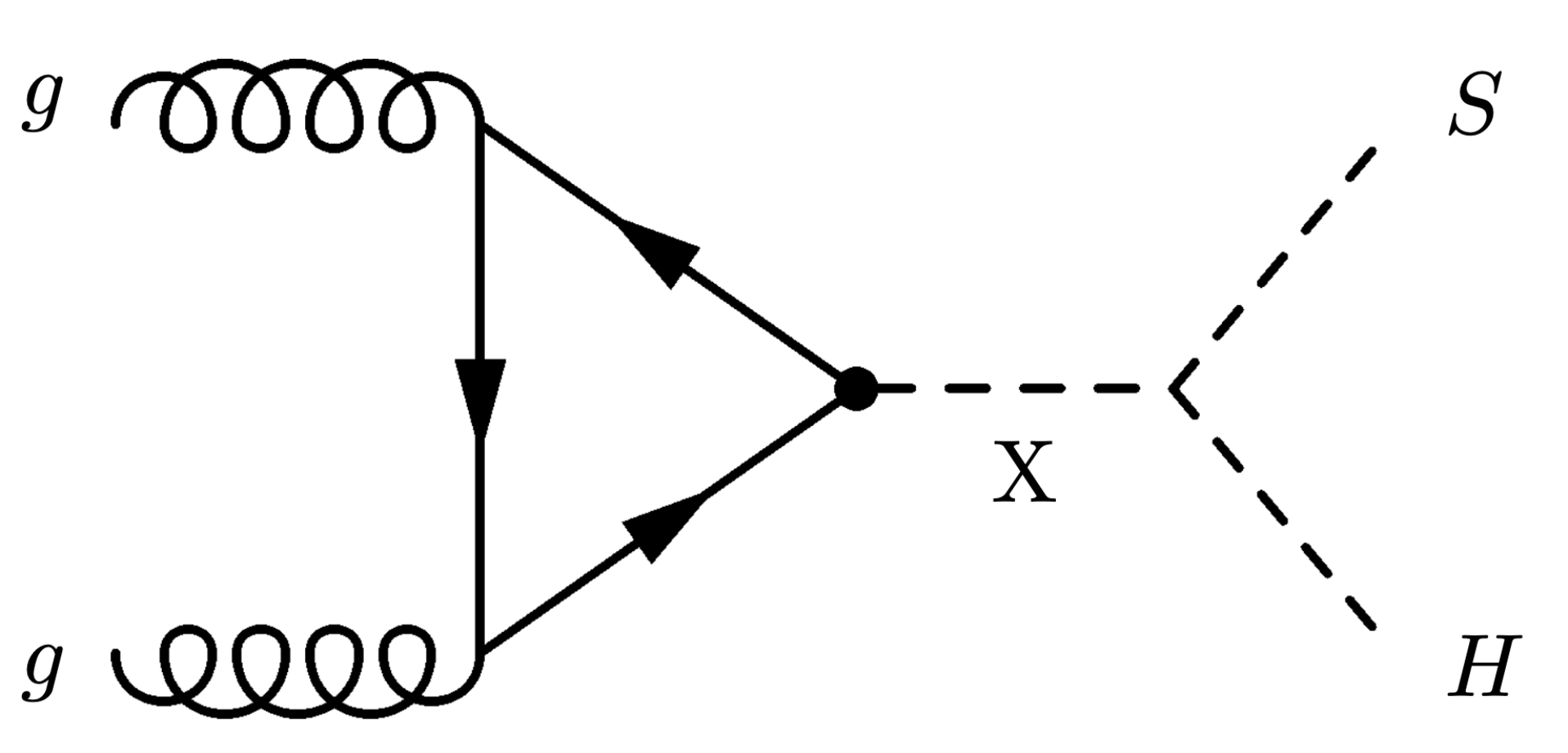}
\caption{Representative diagram that contributes to $X \rightarrow SH$ production via the gluon fusion process.}
\label{fig:X_sh_diagram}
\end{figure}

Many beyond-the-SM (BSM) theories predict the existence of a heavy scalar boson decaying into two Higgs bosons or additional scalars in an extended Higgs sector.  Searches~\cite{HIGG-2013-33,HDBS-2018-58,CMS-HIG-17-030, CMS-HIG-20-014} for these scalars have been published by
the ATLAS and CMS collaborations. The BSM scenarios tested include the two-Higgs-doublet model (2HDM)~\cite{Branco:2hdm2012}, the Minimal Supersymmetric Standard Model (MSSM)~\cite{Fayet:1976et}, and the extensions of the 2HDM model with a new singlet scalar (2HDM+$S$)~\cite{vonBuddenbrock:2016rmr,von_Buddenbrock_2019}.
In the simplest extension of the MSSM, the Next-to-Minimal Supersymmetric Standard Model (NMSSM), an additional gauge singlet is introduced to generate the
$\mu$-term coupling to the superpotential dynamically and this leads to an enriched Higgs sector with two additional neutral Higgs bosons~\cite{Ellwanger_2010, Robens:2019kga}.
The BSM models hypothesize the existence of a new scalar singlet, $S$, in the processes $X\to SH, \,SS$
where $X$ is a heavy CP-even scalar boson produced predominantly through the gluon--gluon fusion process (ggF)~\cite{Muhlleitner:2017dkd}.
The decay of the singlet scalar $S$ is assumed to have the same relative couplings as a SM-Higgs boson of the specified mass. It can be produced via decay mode $X\to SH$ as shown in Figure~\ref{fig:X_sh_diagram}.
 
This study focuses on the decay $X\to SH$, assuming $X$ masses range from 500 to 1500 \GeV, while $S$ is assumed to have a mass range from 200 to 500 \GeV
and decays predominantly into $W^{\pm}W^{\mp}$ and $ZZ$  with the SM-Higgs boson-like decay branching ratios~\cite{deFlorian:2016spz}.
The search is performed by selecting events containing two hadronically
decaying $\tau$-lepton candidates ($\tauhad$) from $H\to \tau^+\tau^-$ decays and one or two light leptons ($\ell=e,\,\mu$)
from $S\to WW,\, ZZ$ decays. This signature-based search is the first of its kind and is competitive with other searches using the
final states $VVbb$, $VV\gamma\gamma$, $bb\tau\tau$~\cite{CMS-HIG-22-001}, and $bbbb$~\cite{CMS:2022suh} in the high mass regions. A direct comparison between them is not
possible due to the unknown branching fractions for $S$ decays, but each provides independent constraints on the BSM models.
To improve background rejection, a multivariate technique based on boosted decision trees (BDTs) is used to discriminate between signal and background based on their different kinematic distributions. 
Upper limits are set at the 95\% confidence level (CL) on the
model-dependent cross section $\sigma(pp\to X\to SH)$ and on the model-independent
$\sigma(pp\to X\to SH\to WW\tau\tau)$ and $\sigma(pp\to X\to SH\to ZZ\tau\tau)$ cross sections.


\section{ATLAS detector}
 
The ATLAS detector~\cite{PERF-2007-01} at the LHC covers almost the entire solid angle around the collision point.\footnote{ATLAS
uses a right-handed coordinate system with its origin at the nominal interaction point (IP) in the
centre of the detector.
The $x$-axis points from
the IP to the centre of the LHC ring, 
the $y$-axis points upward,
and the $z$-axis coincides with the axis of the beam pipe.
Cylindrical coordinates ($r$,$\phi$) are used
in the transverse plane, $\phi$ being the azimuthal angle around the beam pipe. The pseudorapidity is defined in
terms of the polar angle $\theta$ as $\eta = - \ln \tan(\theta/2)$.
Angular separation is measured in units of $\Delta R\equiv \sqrt{(\Delta\eta)^2+(\Delta\phi)^2}$.}
It consists of an inner tracking detector surrounded by a thin superconducting solenoid that produces a
2~T axial magnetic field, electromagnetic and hadronic calorimeters, and a muon spectrometer incorporating three large toroid magnet assemblies with eight coils each. The inner detector contains a high-granularity silicon pixel detector, including the 
insertable B-layer~\cite{IBL1,IBL2,Abbott:2018ikt} added as a new innermost layer in 2014, and a silicon microstrip tracker.  Together they enable the precise reconstruction of tracks of charged particles in the pseudorapidity range $|\eta|<2.5$.
The inner detector also includes a transition radiation tracker that provides tracking and electron identification for $|\eta|<2.0$.
The calorimeter system covers the pseudorapidity range $|\eta| < 4.9$. Within the region $|\eta|< 3.2$, electromagnetic calorimetry is provided by barrel and endcap high-granularity lead/liquid-argon (LAr) sampling calorimeters, with an additional thin LAr presampler covering $|\eta| < 1.8$ to correct for energy loss in material upstream of the calorimeters. Hadronic calorimetry is provided by 
a steel/scintillator-tile calorimeter, segmented into three barrel structures in the region $|\eta| < 1.7$, and two copper/LAr hadronic endcap calorimeters.
The solid angle coverage is completed with forward copper/LAr and tungsten/LAr calorimeter modules optimised for electromagnetic and hadronic measurements, respectively.
The calorimeters are surrounded by a muon spectrometer in a magnetic field provided by air-core toroid magnets with a bending integral of about \SI{2.5}{\tesla\metre} in the barrel and up to \SI{6.0}{\tesla\metre} in the endcaps.
The muon spectrometer measures the trajectories of muons in the region $|\eta|<2.7$ using multiple layers of high-precision tracking chambers, and it is instrumented with separate trigger chambers that cover $|\eta|<2.4$. A two-level trigger system~\cite{Aaboud:2016leb}, consisting of a hardware-based level-1 trigger followed by a software-based high-level trigger, is used to reduce the event rate to a maximum of around \SI{1}{\kHz} for offline storage.
An extensive software suite~\cite{ATL-SOFT-PUB-2021-001} is used in the reconstruction and analysis of real and simulated data, in detector operations, and in the trigger and data acquisition systems of the experiment.


\section{Data and Monte Carlo Simulation samples}
\label{sec:dataset}
The search is based on a data sample of $pp$ collisions at $\sqrt{s}=13~\tev$ with 25~ns bunch spacing collected from 2015 to 2018, corresponding to an integrated luminosity of $140~\ifb$.
Only events recorded with a single-electron trigger, a single-muon trigger, or a dilepton trigger~\cite{TRIG-2018-05,TRIG-2018-01,id_trigger,l1topo_trigger} under stable beam conditions
and for which all detector subsystems were operational are considered for analysis. The average number of $pp$ interactions per bunch crossing in this dataset ranges from 8 to 45, with a mean of 24.
 
Events produced in the generic 2HDM+$S$ model are considered as signals, where the kinematic parameters are model-independent, and the results can be interpreted in other BSM models.
Events are generated to
leading-order (LO) accuracy with \PYTHIA{8.2}~\cite{Sjostrand:2014zea} for matrix element calculation and use the \textsc{NNPDF2.3 LO} set of parton distribution functions (PDF)~\cite{Ball:2014uwa}. Parton showering (PS) and hadronisation are also simulated using the \PYTHIA{8.2} generator with the A14 tune~\cite{ATL-PHYS-PUB-2014-021} and using the same \textsc{NNPDF2.3 LO} PDF set. The detector response is simulated using AltFastII~\cite{ATL-PHYS-PUB-2010-013}, which uses a fast, parameterised simulation of the calorimeter response. Two scalars, $X$ and $S$, are assumed to have a narrow width relative to the experimental resolution. The heavier boson $X$ is constrained to decay only to $S$ and $H$, and $S$ decays to a pair of $W$ or $Z$ bosons. The $S$ decays to SM particles are assumed to be the same as those of a SM Higgs boson of the specified mass. To suppress QCD multijet backgrounds, only semileptonic and fully leptonic decays of $WW$ pairs are considered in the analysis. For $ZZ$ decays, the two-lepton final state is considered, with one Z decaying into two light leptons and the other into a pair of neutrinos or jets.
The SM Higgs boson from the $X$ decay is required to decay into a pair of $\tau$-leptons. Thus three dedicated samples are generated independently by forcing $S\to WW$ or $S\to ZZ$ decays in the following final
states: $\tau\tau+WW(\ell\nu q\bar{q}')$, $\tau\tau+WW(\ell\nu\ell\nu)$ and $\tau\tau+Z(\ell\ell)Z(q\bar{q},\,\nu\nu)$. 
The choice of $m_S$ considered is set to 200~\GeV to enable $S$ decay to on-shell $W$ or $Z$ boson pairs and up to 500~\GeV to have a narrow width approximation. The background contribution
limits the lower bound of  $m_X$ (500~\GeV) while the upper bound of  $m_X$ (1500~\GeV) is required to be in the resolvable $H\to \tau\tau$ decay region.
Eighteen mass points corresponding to various combinations of $m_X$ and $m_S$ values for each of the three final states are generated to cover the most interesting phase space, and are enumerated in grids (in \GeV) of $m_X$ = [500, 750, 1000, 1250, 1500] for $m_S$ = [200, 300] and $m_X$ = [750, 1000, 1250, 1500] for $m_S$ = [400, 500].
Simulated SM non-resonant $HH$ from the ggF process is also generated, for the systematic studies of the $X\to SH$ signals, using the \textsc{Powheg-Box} v2
generator at the next-leading order (NLO)~\cite{Re:2010bp,Alioli:2009je} with the \textsc{NNPDF2.3 LO} PDF set and interfaced with \PYTHIA{8.2} for
PS and hadronisation.
 
Monte Carlo (MC) simulation samples were produced for the different background processes. The dominant background is from falsely identified tau leptons originating from jets. The irreducible background with a real tau-lepton contribution is dominated by diboson processes that produce $\tau\tau$ pairs.
The effects of additional $pp$ collisions in the same or a nearby bunch crossing (pile-up) is modelled using events from minimum-bias interactions generated with \textsc{Pythia8.1}~\cite{Sjostrand:2007gs} using the \textsc{NNPDF2.3 LO} set of PDFs and the A3 set of tuned parameters~\cite{ATL-PHYS-PUB-2016-017},
and overlaid onto the simulated hard-scatter events according to the luminosity profile of the recorded data.
The generated events were processed through a detailed simulation~\cite{Aad:2010ah} of the ATLAS detector geometry and response
using \textsc{Geant4}~\cite{Agostinelli:2002hh} and processed with the same reconstruction software as the data.
Corrections were applied to the simulated events so that the selection efficiencies, energy scales, and energy resolutions of particle candidates match those determined from data control samples.
The simulated samples are normalised to their SM cross sections, computed to the highest available order in perturbation theory.
 
The \ttW and $\ttdy$ samples are generated using the \textsc{Madgraph5\_aMC@NLO} v2.6.2~\cite{Alwall:2014hca} generator, which provides matrix elements at NLO in the strong
coupling constant $\alpha_{S}$ with the \textsc{NNPDF3.0 NLO} PDF set~\cite{Ball:2014uwa}. The functional forms of the renormalisation and
factorisation scales are set to $\mu_r = \mu_f = m_{\mathrm{T}} / 4$ where $m_{\mathrm{T}}$ is defined as the scalar sum of the transverse masses $\sqrt{m^2 + p_{\mathrm{T}}^2}$ of
the particles generated from the matrix element calculation. The events are interfaced with \textsc{Pythia8.2}~\cite{Sjostrand:2014zea} for the parton shower (PS) and hadronisation, using the A14 set of tuned parameters and the \textsc{NNPDF2.3 LO} PDF set.

The $\tth$ process is obtained from
the \textsc{Powheg-Box} v2.0 generator at NLO
and interfaced with \textsc{Pythia8.2} for the parton showering and fragmentation with the A14 tune.
This sample uses the \textsc{NNPDF3.0 NLO} PDF set.
 
Higgs boson production in association with a vector boson ($WH$ or $ZH$, collectively referred to as $VH$) was simulated using \textsc{Powheg-Box} v2 with the \textsc{NNPDF3.0 NLO} PDF set and interfaced
with \textsc{Pythia8.2} for PS and non-perturbative effects. Separate $W^{+}H$, $W^{-}H$, $qq\to ZH$ and $gg\to ZH$ samples are produced. The cross section is
normalised at the next-to-next-to-leading order (NNLO) in QCD with NLO electroweak corrections for \(q\bar{q} \to VH\) and at NLO
and next-to-leading-logarithm accuracy in QCD for \(gg \to ZH\)~\cite{Altenkamp:2012sx,Brein:2012ne}.

The $\ttbar$ events are generated with \textsc{Powheg-Box} v2.0 and interfaced with \textsc{Pythia8.2} for the parton showering and fragmentation with the A14 tune. The PDF set of \textsc{NNPDF3.0 NLO} is used.
The single top-quark events are simulated with \textsc{Powheg-Box} v2.0 with the \textsc{NNPDF3.0 NLO} PDF set and
interfaced with \textsc{Pythia8.2}, where the interference
between $Wt$ ad $\ttbar$ production is handled with the Diagram Removal (DR) procedure~\cite{Frixione:2008yi}.

A dedicated $\ttbar$ sample including rare $t \to Wb \gamma^*(\to l^+l^-)$ radiative decays, $\ttbar \to W^+bW^-\bar{b}l^+l^-$,
is generated using a \textsc{Madgraph5\_aMC@LO} matrix element requiring $m(l^+l^-) > 1~\gev$ and interfaced with \textsc{Pythia8.2}.
In this sample, the photon can be radiated from the top quark, the $W$ boson,
or the $b$-quark. Both the $\ttdy$ and $\ttbar \to W^+bW^-\bar{b}l^+l^-$ samples are combined and together form the ``$\ttZ$ (high mass)'' sample.
The contribution from internal photon conversions ($\gamma^* \to l^+l^-$) with $m(l^+l^-) < 1~\gev$ is
modelled with \textsc{Sherpa} 2.2.11~\cite{Bothmann:2019yzt} through QED multiphoton radiation via the PS in an inclusive $\ttbar$
sample and is referred to as ``$\ttbar\gamma^*$ (low mass)'' sample.
 
Diboson backgrounds are generated and normalised using the cross sections computed by \textsc{Sherpa} 2.2.2~\cite{Bothmann:2019yzt}
using the \textsc{NNPDF3.0 NNLO} PDF set and showered using Sherpa's default setting.
The cross section is computed at NNLO accuracy in QCD and includes the electroweak (EW) corrections at NLO accuracy.
The processes of $W$ and $Z$ production associated with jets ($V$+jets) are simulated with \textsc{Sherpa} 2.2.1~\cite{Bothmann:2019yzt} using the \textsc{NNPDF3.0 NNLO} PDF and showered by the \textsc{Sherpa} built-in
implementation using matrix elements for up to two additional jets at NLO and up to four additional jets at LO.
The cross section used to normalise the simulation
is calculated at NNLO accuracy in QCD and includes EW corrections at NLO accuracy. Low-mass Drell--Yan $m_{\ell\ell}$ production is also included.
 
The MC background samples are divided into five categories: data-driven fake \tauh, $t\bar{t}H$, $t\bar{t}V$, diboson, and 'others,' which includes $t\bar t$, $V$+jets, low-mass Drell--Yan, $VH$, $Wt$, and other small backgrounds.


\section{Object reconstruction}
\label{sec:objects}
 
Events are required to have at least one primary vertex with a minimum of two associated tracks, each with a transverse momentum ($\pt$) larger than $500~\mev$. The primary vertex of an event is defined as the vertex with the highest scalar sum of squared transverse momenta of the associated tracks~\cite{ATL-PHYS-PUB-2015-026}.
 
Electrons are reconstructed by matching tracks reconstructed in the inner detector (ID) to topological clusters of energy deposits in the electromagnetic calorimeter. They are required to have $\pt>10$ \GeV\ and $|\eta|<2.47$, and to be outside the transition region between the barrel and endcap calorimeters, $1.37<|\eta|<1.52$. Electron candidates must satisfy a \textit{Loose} likelihood-based identification requirement~\cite{EGAM-2018-01}. Additional requirements on the associated track $\pt$ and the ratio of the electron’s calorimeter energy and the track momentum are applied to electrons to suppress photon conversions. Electrons with incorrect charge assignments (arising from asymmetric photon conversions) are rejected using a BDT discriminant based on calorimeter and tracking quantities. An efficiency of $95\%$ for electrons with correct charge assignment is obtained with a rejection factor of $\sim 17$ for electrons with incorrect charge assignment~\cite{EGAM-2018-01, PERF-2017-01}.
 
Muon candidates are reconstructed from tracks in the muon spectrometer (MS) matched to tracks found in the ID. They are required to have $\pt>10$ \GeV\ and $|\eta|<2.5$ and to satisfy \textit{Loose} identification requirements~\cite{PERF-2015-10} based on the numbers of hits in the different ID and MS subsystems and on the significance of the charge-to-momentum ratio $q/p$.
 
Electrons (muons) are required to have associated tracks satisfying $|d_{0}|/\sigma_{d_{0}} < 5 (3)$ and $|z_{0} \mathrm{sin}\theta| < 0.5$ mm, where $d_{0}$ is the transverse impact parameter relative to the beam line, and $z_{0}$ is the longitudinal impact parameter relative to the primary vertex.

Jets are reconstructed using the anti-$k_t$ jet clustering algorithm~\cite{Cacciari:2008gp,Cacciari:2011ma} with a radius parameter of $R=0.4$, applied to topological energy clusters~\cite{Aad:2016upy, ATL-PHYS-PUB-2015-036} and charged-particle tracks, processed using a particle-flow algorithm~\cite{PERF-2015-09}. 
The reconstructed jets are corrected by the application of jet energy scale (JES) and resolution (JER) calibrations derived from simulation and in situ corrections based on $\sqrt{s}=13~\tev$ data~\cite{Aaboud:2017jcu, JETM-2018-05}. Only jet candidates with $\pt > 25~\gev$ and $|\eta| < 2.5$ are considered. Jets with $\pt<60~\gev$ and $|\eta| < 2.4$ originating from pile-up are suppressed by the use of the jet vertex tagger (JVT)~\cite{Aad:2015ina}. Quality criteria are imposed to reject events containing jets from non-collision backgrounds and calorimeter noise~\cite{ATLAS-CONF-2015-029}.
 
Jets containing $b$-hadrons ($b$-jets) are identified with the DL1r tagger algorithm~\cite{FTAG-2018-01,ATL-PHYS-PUB-2017-013}, a multivariate discriminant based on inputs such as the impact parameters of displaced tracks and the reconstructed secondary vertices in the jet. Working points are defined with different target efficiencies for tagging $b$-jets from an inclusive $\ttbar$ sample. This analysis uses the 77\% $b$-tagging working point to veto $b$-jets.
 
The reconstruction of the hadronically decaying $\tau$-lepton candidates (\tauhad) is seeded by jets reconstructed with an anti-$k_t$ algorithm using calibrated topological clusters with a radius parameter of $R=0.4$~\cite{Aad:2016upy, ATLAS-CONF-2017-029}. They are required to have $\pt > 20$ \GeV\ and $|\eta|<2.5$, excluding the transition region between the barrel and endcap calorimeters ($1.37<|\eta|<1.52$). The \tauhad candidates are required to have one or three associated tracks and a total charge of $\pm$1. An identification algorithm based on a recurrent neural network (RNN)~\cite{ATL-PHYS-PUB-2019-033} discriminates the visible decay products of the \tauhad candidates from jets initiated by quarks or gluons. The \tauhad candidates are required to satisfy the RNN \textit{Medium} identification working point in the signal regions, corresponding to an efficiency of $75\%$ ($60\%$) for candidates with one (three) associated track(s). The efficiency is optimized to be flat as a function of the \tauhad \pt and pile-up. A dedicated BDT discriminant is used to reject electrons misidentified as hadronic tau decays. To suppress fake \tauhad candidates from muons, the candidates that overlap with low-$\pt$ reconstructed muons are vetoed. Efficiency scale factors for the reconstruction, identification, and electron BDT rejection are applied to \tauhad candidates in simulation~\cite{HIGG-2019-09}.
 
Anti-\tauhad candidates are defined to estimate the background from jets misidentified as \tauhad, as described in Section~\ref{sec:background}. They are required to satisfy the RNN \textit{Very Loose} identification working point
but fail the nominal RNN requirement applied to the \tauhad candidates.
 
An overlap removal procedure is applied when two reconstructed objects are close in $\Delta R$ to avoid double-counting of objects.
 
Isolation criteria are applied to electrons and muons to suppress contributions from semileptonic decays of heavy-flavour hadrons or jets misidentified as leptons, collectively referred as non-prompt leptons, after the overlap removal procedure. A BDT discriminant is trained based on isolation and $b$-tagging variables and is referred to as the non-prompt lepton BDT~\cite{HIGG-2017-02}.  
The light leptons must satisfy the \textit{Loose} BDT working point to reject non-prompt leptons.
 
The missing transverse momentum $\mpt$ (with magnitude $\met$) is defined as the negative vector sum of the $\vec{p}_{\mathrm{T}}$ 
of all selected and calibrated objects in the event, including a term to account for soft particles that are not associated with any of the selected objects. This soft term is calculated from ID tracks matched to the selected primary vertex to make it more resilient to contamination from pile-up interactions~\cite{Aaboud:2018tkc}.


\section{Event selection}
\label{sec:selection}
In this search three signal regions (SR) are considered, referred to as $\wwlditau$, $\wwllditau$, and $\zzllditau$ channels. Events in the $\wwllditau$ and $\zzllditau$ channels contain exactly two light leptons with opposite-sign electric charges, while events in the $\wwlditau$ channel contain exactly one light lepton.
 
Selected events must satisfy either the un-prescaled single lepton (SL) or dilepton (DL) triggers. The SL triggers are required in the $\wwlditau$ channel
while a logical ``or'' between SL and DL triggers is required for the $\wwllditau$ and $\zzllditau$ channels. The offline reconstructed light leptons in the event must be matched to the trigger-level objects that caused the event to pass the trigger(s).
The reconstructed light lepton $\pt$ must exceed the threshold set for the online trigger object by at least 1~\GeV\ to avoid threshold biases~\cite{HIGG-2017-02}. 
 
A $Z$ candidate must be present in the $\zzllditau$ channel. It is defined by two leptons of the same flavour and opposite charge with an invariant mass that is consistent with the $Z$ boson mass, $|m_{\ell\ell}-m_{Z}|<10$ \GeV.
In the $\wwllditau$ channel the dilepton invariant mass must satisfy $m_{\ell\ell} > 12$ \GeV\ and $Z\to\ell\ell$ candidates are vetoed.
 
For each channel, exactly two RNN medium $\tauhad$ candidates of opposite charge with $\pt > 20$ \GeV\ are required. The angular distance between the two $\tauhad$ candidates is required to satisfy $\Delta R_{(\tau_0,\tau_1)} \leq 2$ to suppress $V$+jets background events.

The event selection for each channel is summarised in Table~\ref{tab:sr_definition}. A multivariate (MVA) technique is employed in all channels to enhance the sensitivity by discriminating signals from the SM
background processes. The description of the MVA strategy is given in Section~\ref{sec:bdt}.
 
\begin{table}[!h]
\caption{Selection criteria applied in the $\wwlditau$, $\wwllditau$ and $\zzllditau$ signal regions
where $m_{Z}$=91.2 \GeV\,is used.}
\label{tab:sr_definition}
\begin{center}
\begin{tabular}{l|l}\toprule\toprule
Channels            & Selections \\
\midrule
\wwlditau           & exactly one light lepton (electron or muon): $\pt>27$ \GeV,\, $|\eta|<2.5$\\
& exactly two RNN medium $~\tauhad$ with opposite-sign: $\pt>20$ \GeV,\, $|\eta|<2.5$ \\
& $\Delta R$ between two $\tauhad$ candidates: $\Delta R_{(\tau_0,\tau_1)} \leq 2$ \\
& number of jets and $b$-jets: \njets $\geq$ 2 and \nbjets == 0 \\
\midrule
\wwllditau          & exactly two light leptons with opposite-sign: $\pt>10$ \GeV,\, $|\eta|<2.5$\\
& exactly two RNN medium $~\tauhad$ with opposite-sign: $\pt>20$ \GeV,\, $|\eta|<2.5$\\
& invariant dilepton mass: $m_{\ell\ell} > 12$ \GeV \\
& Z-veto ($|m_{\ell\ell}-m_{Z}|>10$ \GeV) for same-flavour leptons\\
& $\Delta R_{(\tau_0,\tau_1)} \leq 2$ \\
& \nbjets == 0 \\
\midrule
\zzllditau          & exactly two same-flavour light leptons with opposite-sign: $\pt>10$ \GeV,\, $|\eta|<2.5$\\
& exactly two RNN medium $~\tauhad$ with opposite-sign: $\pt>20$ \GeV,\, $|\eta|<2.5$\\
& Z-peak selection ($|m_{\ell\ell}-m_{Z}|<10$ \GeV)\\
& $\Delta R_{(\tau_0,\tau_1)} \leq 2$ \\
& \nbjets == 0 \\
\bottomrule\bottomrule
\end{tabular}
\end{center}
\end{table}
 
Same-sign validation regions (VR) are also selected using the same event selection as in the signal regions, except that the two $\tauhad$ candidates must have the same charge (SS). In this VR, the signal contribution is negligible, and the background modelling can be checked using data.


\section{Multivariate discriminant}
\label{sec:bdt}
 
Boosted decision trees implemented in the Toolkit for Multivariate Analysis (TMVA)~\cite{TMVA} are used in each SR to improve the separation between signal and background.
In the training process, all signal and background events are divided into independent training and test samples to avoid possible biases.
To simplify the training procedure across the 18 mass hypotheses of $X$ and $S$ particles,
a parameterised BDTG method~\cite{pTMVA} is used. This method trains a BDT using several $m_{X}$ samples
with the same $m_S$ by including their generated $m_X$ mass as an input parameter while the $m_X$ values of background events are
assigned randomly to match the signal population.
The choice made for the parameterisation of the BDT is motivated by the fact that the signal kinematics are driven by the two-body decay $X\to SH$ for the given $m_S$.
The training algorithm incorporates the $m_X$ parameter so the
BDT is parameterised in $m_X$ for given $m_S$.
This procedure is performed separately in three channels ($\wwlditau$, $\wwllditau$, and $\zzllditau$) with
four $m_S$ masses (200, 300, 400, 500 \GeV), so 12 different BDTs are trained.
 
Many potential variables were investigated in each SR separately. These variables are chosen as they are expected to provide good separation
between the background and signal topologies. Figure~\ref{fig:plots_vv2taupostfit} shows the distributions of some highly discriminating
variables described in Table~\ref{tab:bdt_inputs} in each SR. The discrimination of a given variable is quantified by the `separation'
(which measures the degree of overlap between background and signal distributions) and `importance' (which ranks the power of
the variable in the classification of the events) provided by the TMVA package.
The BDT discriminant is trained for each SR starting from an extensive list of variables; then, the least important variables are
removed sequentially, and the BDT is retrained until the area under the receiver operating characteristic (ROC) curve drops by more than 1\%.
The final selected BDT input variables in each SR are summarised in Table~\ref{tab:bdt_inputs}. In the $\zzllditau$ channel, no explicit selection is applied
for the second $Z$ decay. But, it is required to have a sum of large $\met$ from $Z\to\nu\nu$ and jets \pt from $Z\to qq$ decay in the BDT training
to improve the background rejection without further dividing the data.

\begin{table}[htbp]
\caption{Input variables used in the BDT training for the $\wwlditau$, $\wwllditau$, and $\zzllditau$ channels, where ``$\times$'' stands for used and ``-'' stands for not used. The minimum angular distance is evaluated by using all jets in an event.}
\label{tab:bdt_inputs}
\begin{center}
\resizebox{\textwidth}{!}{
\begin{tabular}{l|l|ccc}\toprule\toprule
Variable & Definition &  \makecell{$\ensuremath{WW}$ \\ $1\ell2\ensuremath{\tau_\textrm{had}}$} & \makecell{$\ensuremath{WW}$ \\ $2\ell2\ensuremath{\tau_\textrm{had}}$} & \makecell{$\ensuremath{ZZ}$ \\ $2\ell2\ensuremath{\tau_\textrm{had}}$}\\ \midrule
$m_{\text{X, truth}}$ & generated mass of generated $X$ particle & $\times$& $\times$& $\times$ \\
$\Delta R(\tau\tau,\ell_0)$ & angular distance between the leading lepton and the $\tau\tau$ system & $\times$ & $\times$ & $\times$ \\
\text{min}($\Delta R(\tau\tau,\text{j})$) & minimum angular distance between a jet and the $\tau\tau$ system & $\times$ &-&- \\
$\Delta R(\ell,\ell)$ & angular distance between two leptons & -&$\times$&$\times$ \\
$\Delta \phi(\tau\tau,\met)$ & azimuthal angle between the $\tau\tau$ system and $\met$ & $\times$&$\times$&$\times$ \\
$\met+\Sigma{\pt}(\text{jets})$ & sum of $\met$ momentum and $\pt$ of jets & -&-&$\times$ \\
${\pt}_{\tau0}$ & leading tau-lepton $\pt$ & $\times$&$\times$& $\times$ \\
$m_{\tau\tau}$ &  visible invariant mass of the $\tau\tau$ system & $\times$&$\times$&$\times$ \\
$m_{\ell\ell}$ & invariant mass of the dilepton system & -&$\times$&- \\
\text{min}($\Delta R(\ell,\text{j})$ & minimum angular distance between a jet and the lepton & $\times$&-&- \\
\text{min}($\Delta R(\text{j},\text{j})$) & minimum angular distance between two jets & $\times$&-&- \\
${\pt}_{\tau1}$  & subleading $\tau$-lepton $\pt$ & $\times$&$\times$&$\times$ \\
$m^{W}_{\text{T}}$ & transverse mass calculated from the lepton(s) and $\met$ in the event & $\times$&$\times$&$\times$ \\
dilep\_type & dilepton type: one of $\mu\mu$, $e \mu$, $\mu e$, $ee$ & -&$\times$&- \\
\bottomrule\bottomrule
\end{tabular}
}
\end{center}
\end{table}
 
The BDT modelling in the signal regions is also checked using the data in the SS VR and in the low BDT regions with BDT $<$ 0.1, where the signal contributions are negligible. The data is consistent with the background expectations in both validation regions.
The final observable used to extract the signal contribution is the BDT distribution in each SR.


\section{Background Modelling}
\label{sec:background}
 
The reducible backgrounds arise from events where at least one of the $\tauhad$ candidates originates from a source other than vector boson decay.
A quark or gluon-initiated jet is usually misidentified as a $\tauhad$ candidate in such backgrounds.
These backgrounds with fake \tauhad candidates are estimated using the data-driven fake factor (FF) method~\cite{ATLAS-CONF-2017-029}.
The irreducible backgrounds with a real $\tauhad$ candidate are modelled using simulation.
 
In all channels, the largest contribution to the total fake \tauh background, including leading and subleading \tauh candidates, arises from the \vjet process.
The dominant source of the fake \tauh are light-quark jets. Other types of fakes include gluon-initiated, $b$-, and $c$- jets,
and an electron or muon misidentified as a fake \tauh candidate. The background contribution from non-prompt
light leptons ($\ell=e,\,\mu$) is small in the signal regions and estimated from simulation.
 
The FF method is used to estimate the contribution of fake \tauh background in the $\wwlditau$, $\wwllditau$, and $\zzllditau$ signal regions.
The method uses an extrapolation from a dedicated control region (CR) enriched in fake \tauh candidates to estimate the number of events with fake \tauh in the signal region by rescaling the templates of fake \tauh in the CR with the fake factors that discussed below.
The CR event selection is the same as that used in the signal region, except that one or both \tauh
candidates fail the RNN medium criteria but pass the RNN \textit{Very Loose} (RNN $>$ 0.05) criteria, referred to as anti-taus or fake $\ensuremath{\tau_\textrm{had}}$.
The templates are produced by subtracting the real \tauh contributions from the simulation.
 
The fake factor is a transfer factor defined as the ratio of the number of events with RNN medium \tauh candidate to the number of events with
anti-\tauh candidates. FFs are measured from data in two dedicated control regions.
These control regions are designed to be enriched in \zjet ($Z(\ell\ell)$) and \ttbar ($b$-tagged dilepton) backgrounds, by requiring zero b-tagged jets, in the former case, and one b-tagged jet and a Z boson veto in the latter case.
The contribution of events with a real \tauh candidate from the simulation is at the 2\% level and is subtracted in these control regions.
The measured FFs in the \zjet and \ttbar control regions are consistent within the statistical uncertainties; however, differences between FFs are treated as systematic uncertainties arising from the different jet compositions.
The fake \tauh background estimate is validated in a dedicated validation region for each channel.
An overall 10\% non-closure uncertainty is estimated by taking the ratio of data to the total predicted background in the $\wwlditau$
same-sign validation region.

The final irreducible and data-driven fake \tauh background contributions in the signal regions are summarised in
Table~\ref{tab:FinalevntYieldspostfit_vv2taus}. The data are consistent with the background expectation.
The leading sources of systematic uncertainty in the fake \tauh background before the fit to data come from
the subtraction of
real \tauh contribution ($^{+7.5}_{-12}$\%), the non-closure uncertainty ($\pm$10\%), and the uncertainty due to the fake \tauh background composition differences between the $Z$+jets and \ttbar background control regions ($\pm$2\%).
 
Event kinematic distributions are compared between the data and the predicted background for the visible invariant mass of the di-\tauh system ($m_{\tau\tau}$), the leading \tauh $\pt$, and other variables, after using a background-only fit to the data. Figure~\ref{fig:plots_vv2taupostfit} shows the distributions for the $\wwlditau$, $\wwllditau$ and $\zzllditau$ channels. Good agreement is found between the data and the predicted background.

\begin{table}[htbp]
\caption{Event yields after a background-only fit and data in the three signal regions.
All the backgrounds except the fake \tauh are estimated using the real \tauh contributions from MC. The fake \tauh background contribution is data-driven.
The top-quark, $V$+jets, and small backgrounds are included in the ``Others'' category. The total uncertainties include systematic and statistical contributions. In the $\wwlditau$ channel, the smaller uncertainty in the total background than the fake \tauh is due to the anti-correlations between the nuisance parameters of the fake \tauh and diboson backgrounds obtained in the fit.}
\label{tab:FinalevntYieldspostfit_vv2taus}
\begin{center}
\setlength{\tabcolsep}{12pt}
\begin{tabular}{l|c|c|c}\toprule\toprule
Process  & {$\wwlditau$} & {$\wwllditau$} & {$\zzllditau$} \\
\midrule
$\mathrm{t\bar{t}}\text{H}$ & $2.6\pm0.3$ & $0.50\pm0.06$ & $0.035\pm0.004$ \\
$\mathrm{t\bar{t}}\text{V}$ & $3.4\pm0.4$ & $0.58\pm0.07$ & $0.10\pm0.02$ \\
Others & $15.6\pm3.0$ & $2.1\pm0.4$ & $4.7\pm0.8$ \\
Diboson & $135.1\pm15.5$ & $11.1\pm1.3$ & $55.0\pm6.1$ \\
Fake $\tau_{\mathrm{had}}$ & $312.4\pm21.5$ & $30.1\pm3.7$ & $77.8\pm9.0$ \\ \midrule
Total background & $468.6\pm19.3$ & $44.0\pm3.9$ & $138.0\pm9.1$ \\ \midrule
Data  & $464$ & $40$ & $138$ \\
\bottomrule\bottomrule
\end{tabular}
\end{center}
\end{table}

\begin{figure}[htbp]
\centering
\begin{tabular}{@{}ccc@{}}
\includegraphics[width=0.28\textwidth]{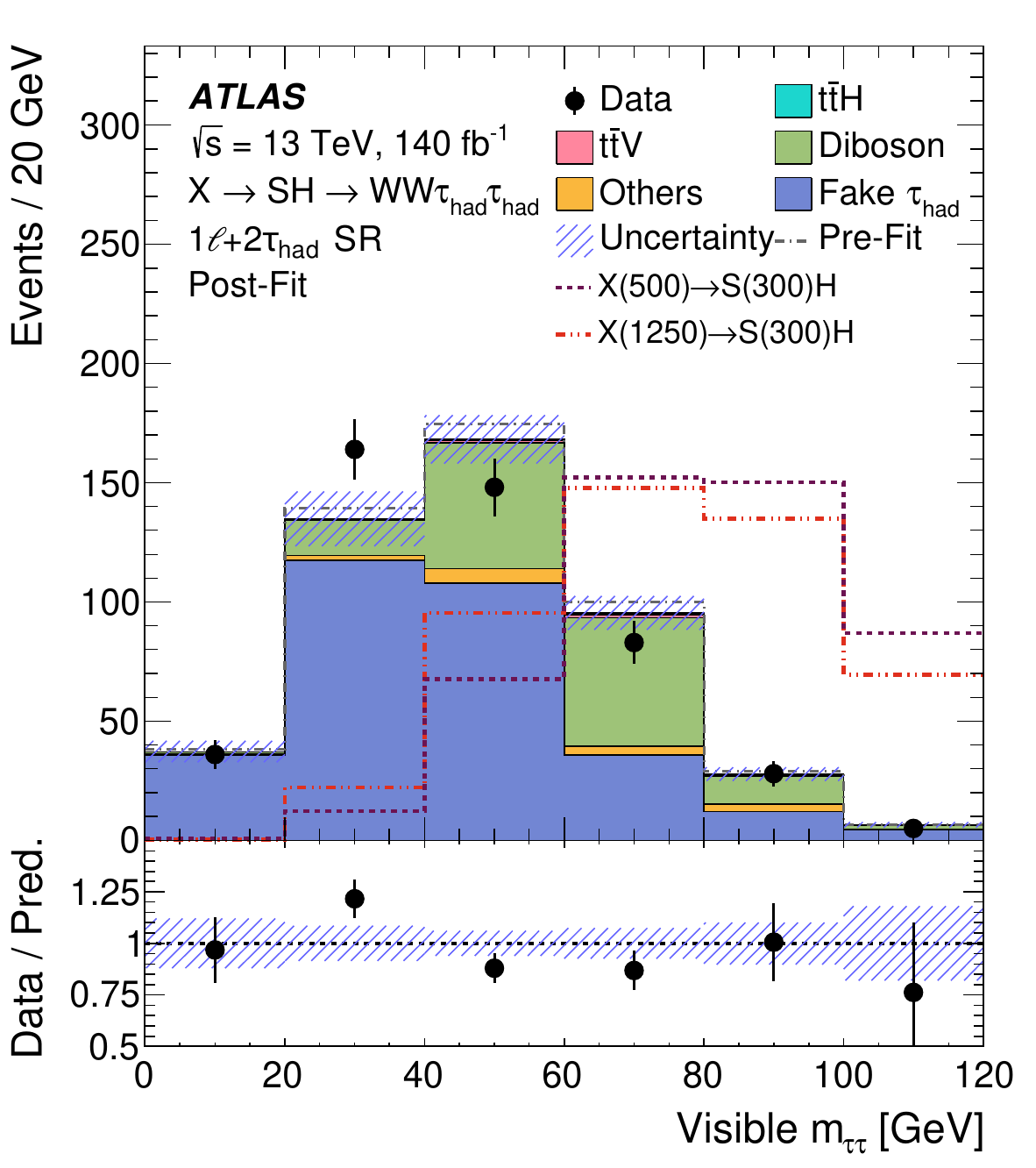} &
\includegraphics[width=0.28\textwidth]{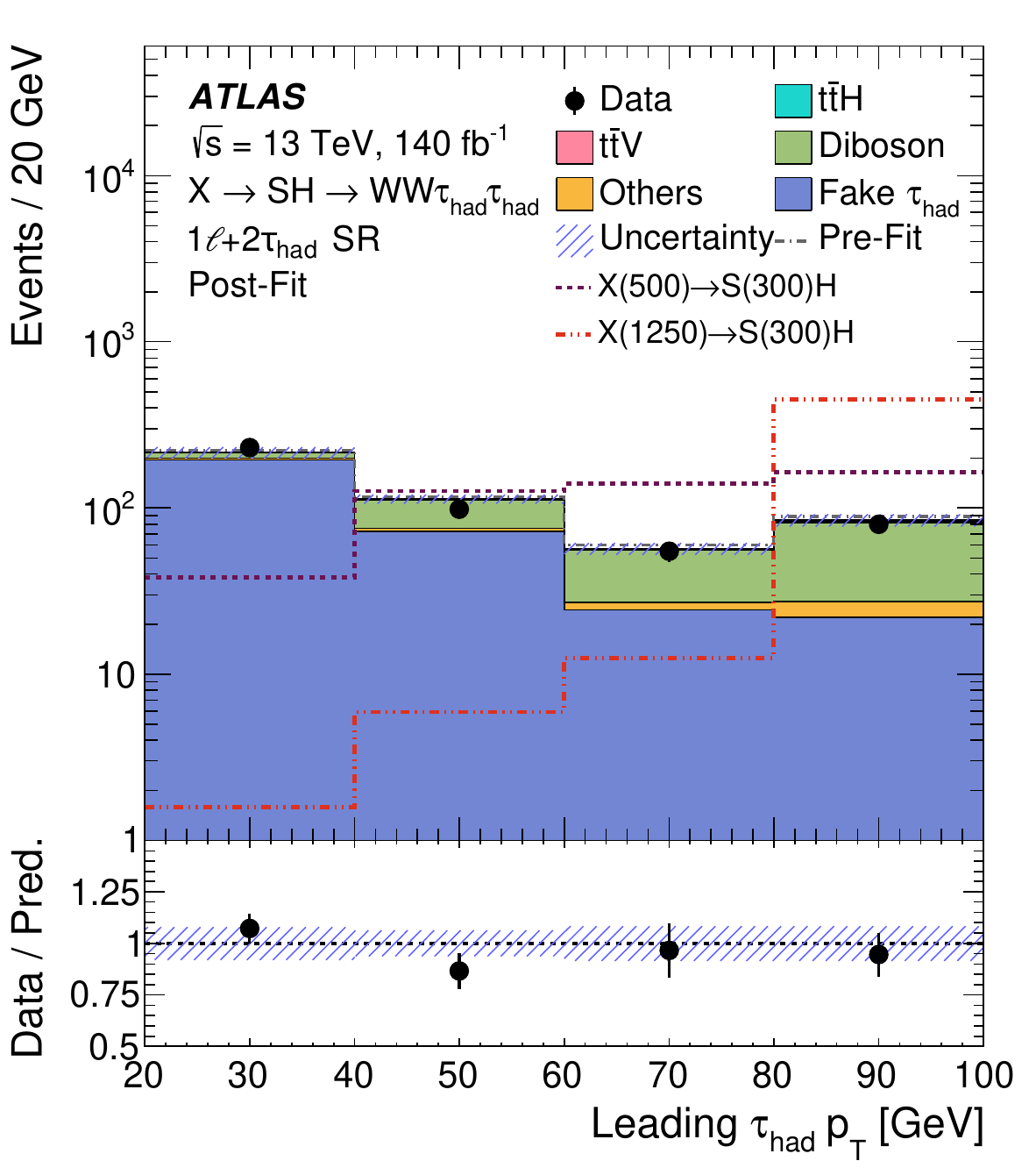} &
\includegraphics[width=0.28\textwidth]{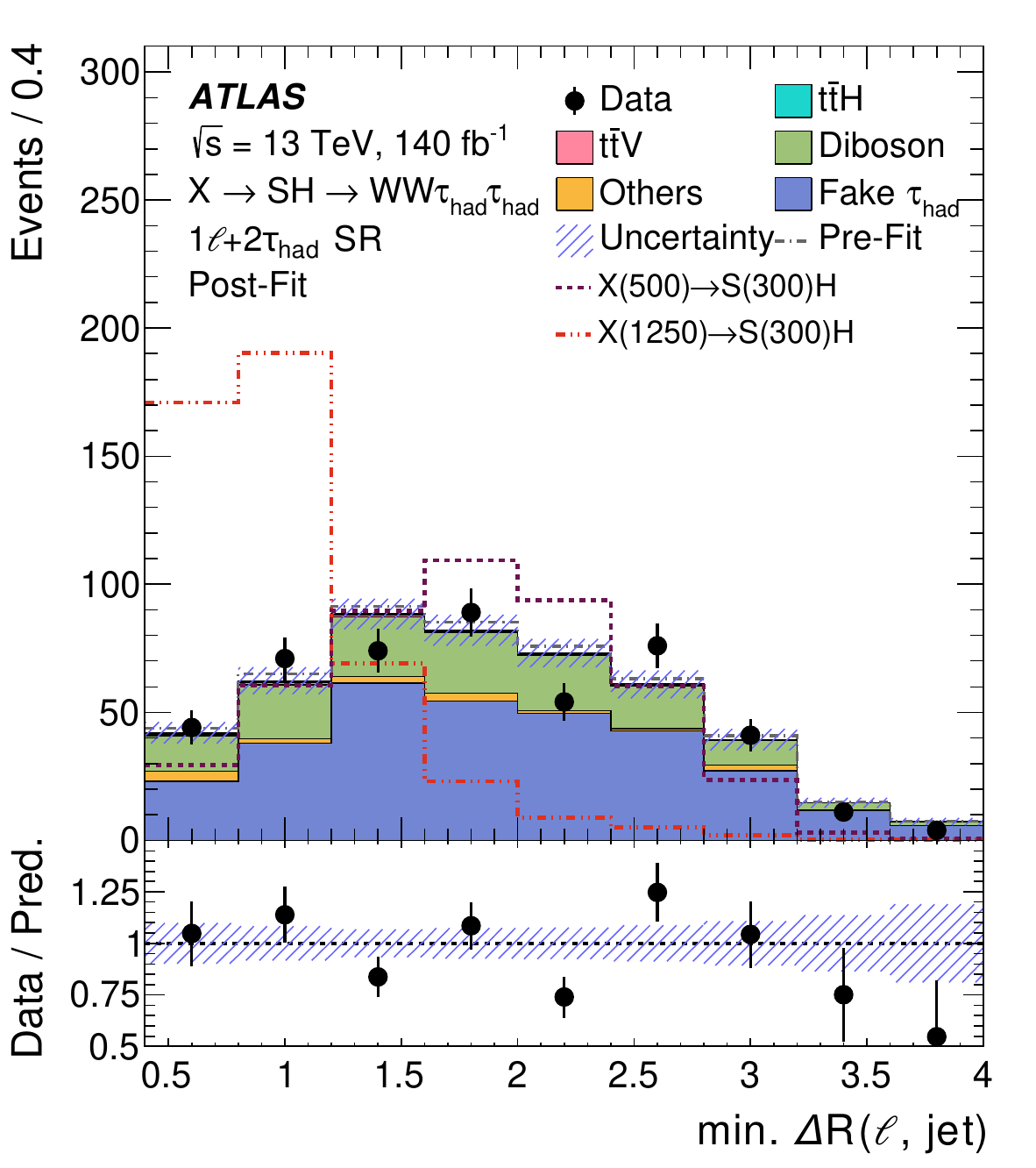} \\
(a) & (b) & (c) \\
\includegraphics[width=0.28\textwidth]{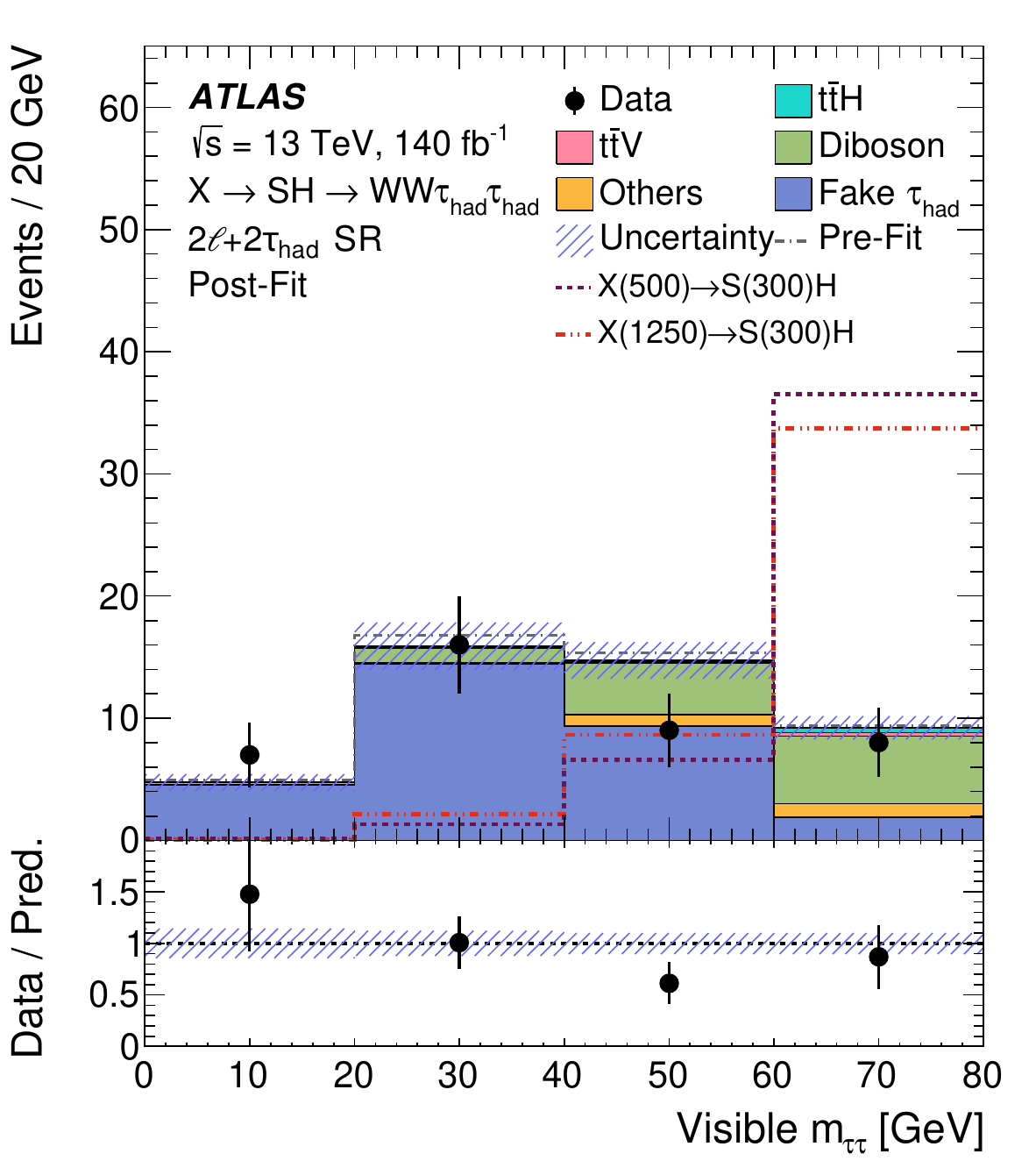}&
\includegraphics[width=0.28\textwidth]{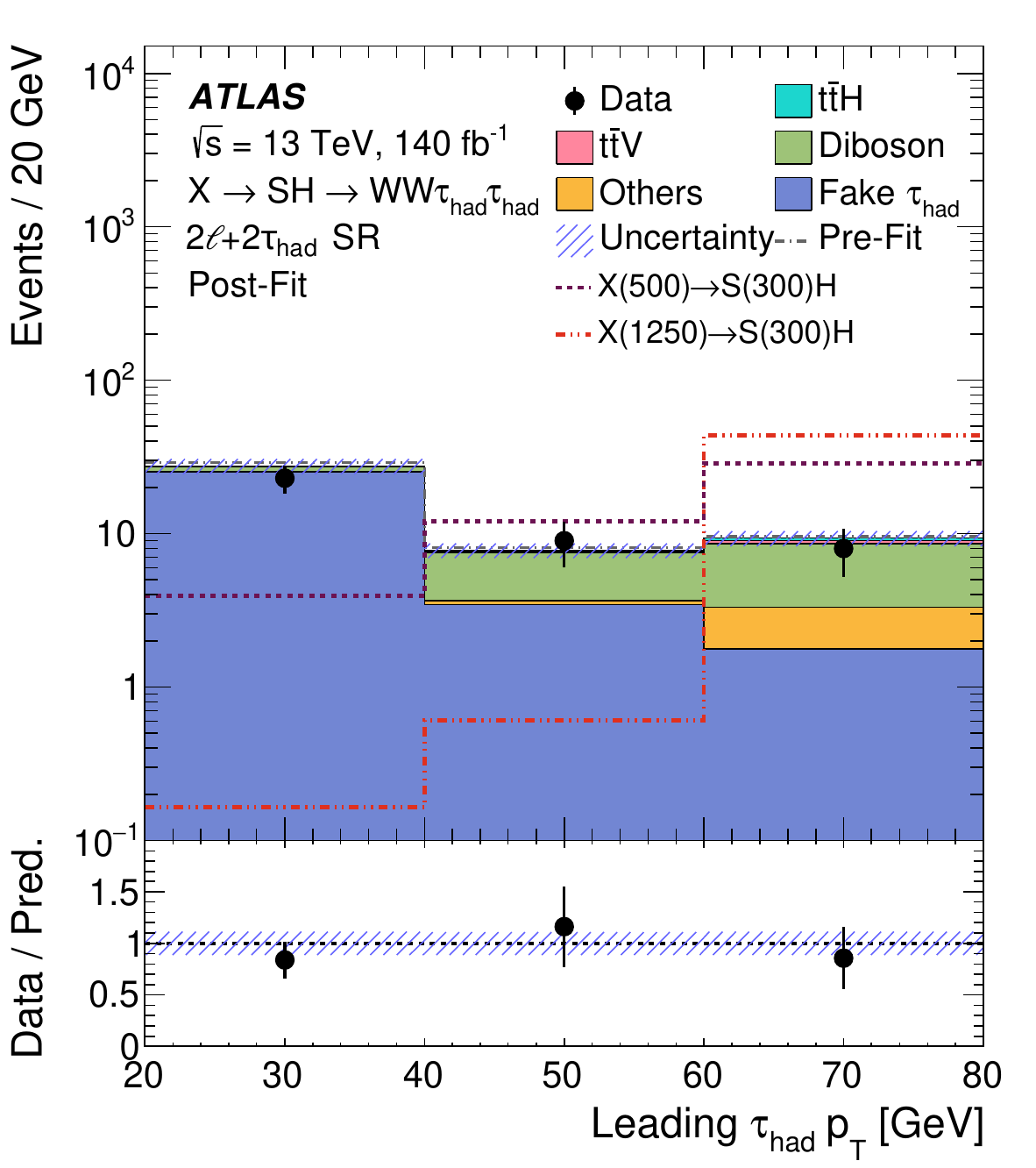} &
\includegraphics[width=0.28\textwidth]{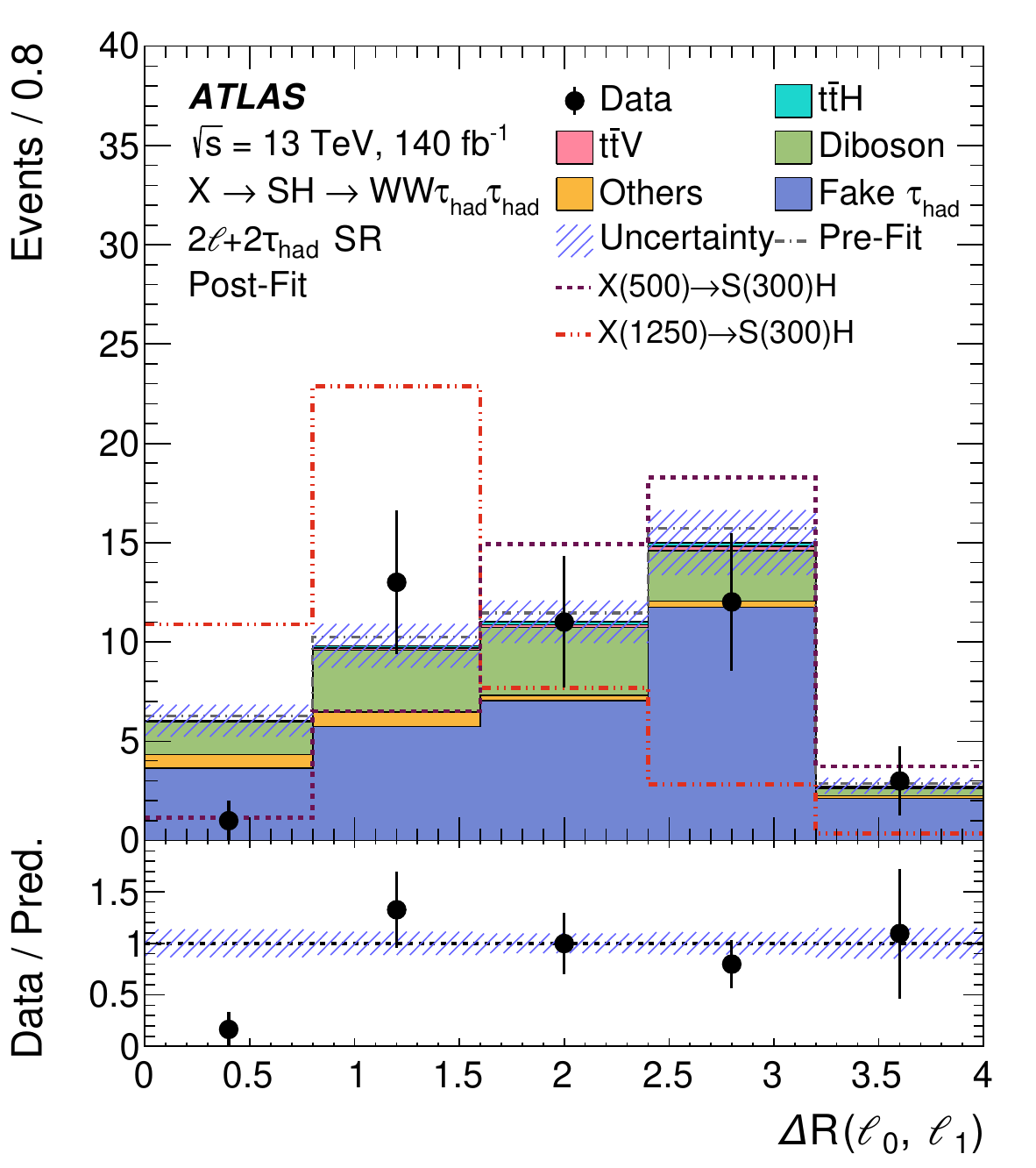} \\
(d) & (e) & (f)\\
\includegraphics[width=0.28\textwidth]{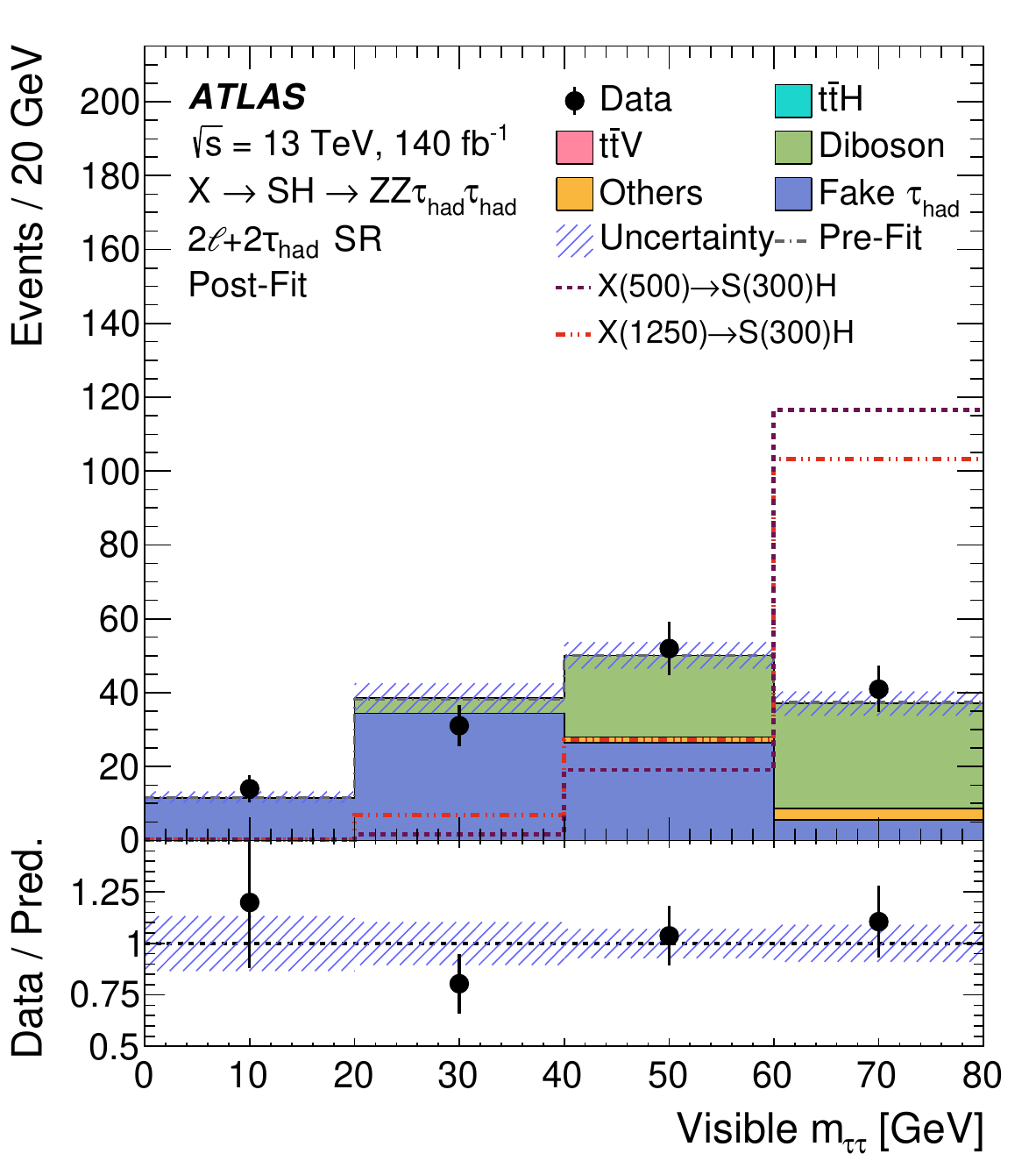}&
\includegraphics[width=0.28\textwidth]{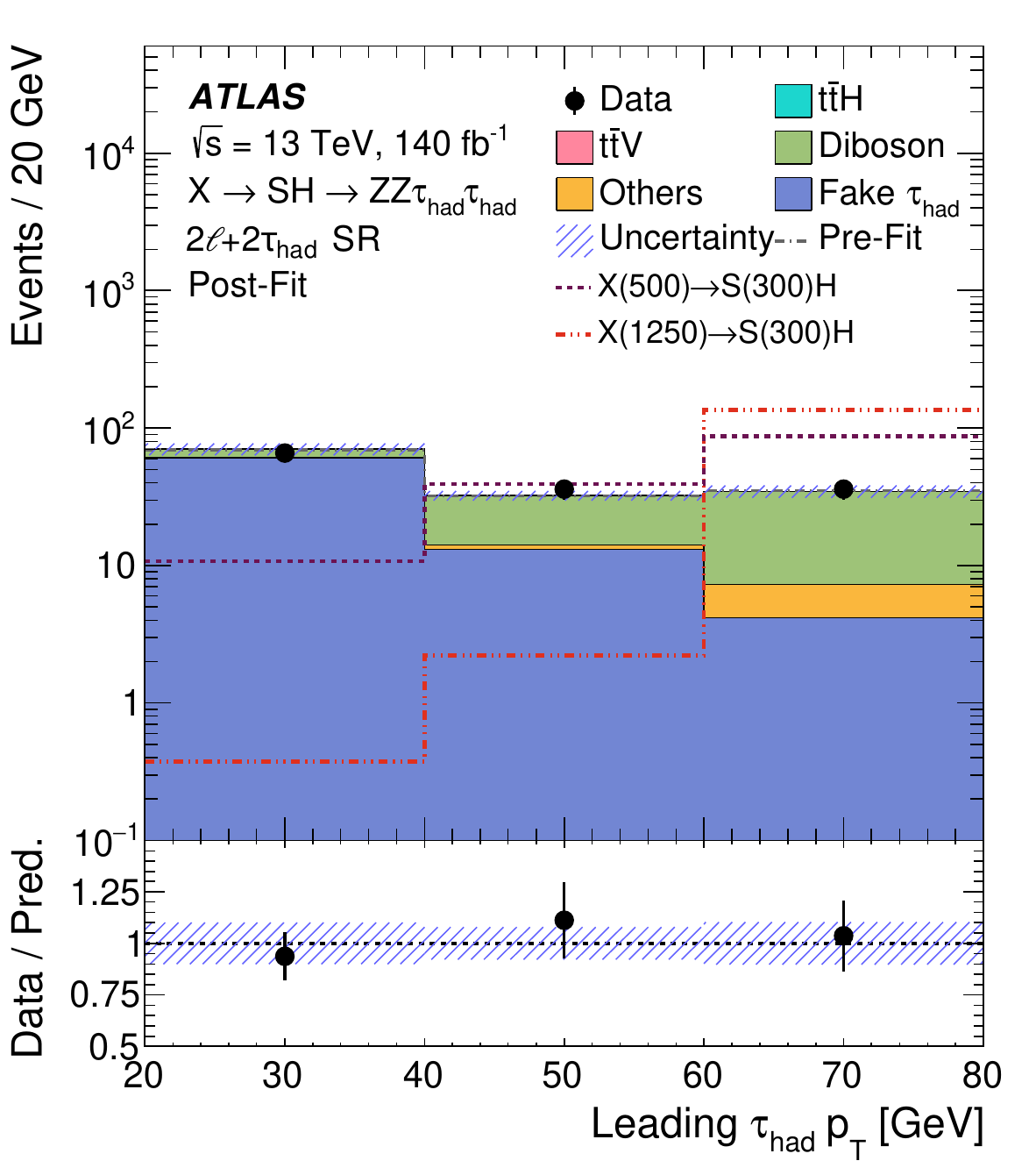} &
\includegraphics[width=0.28\textwidth]{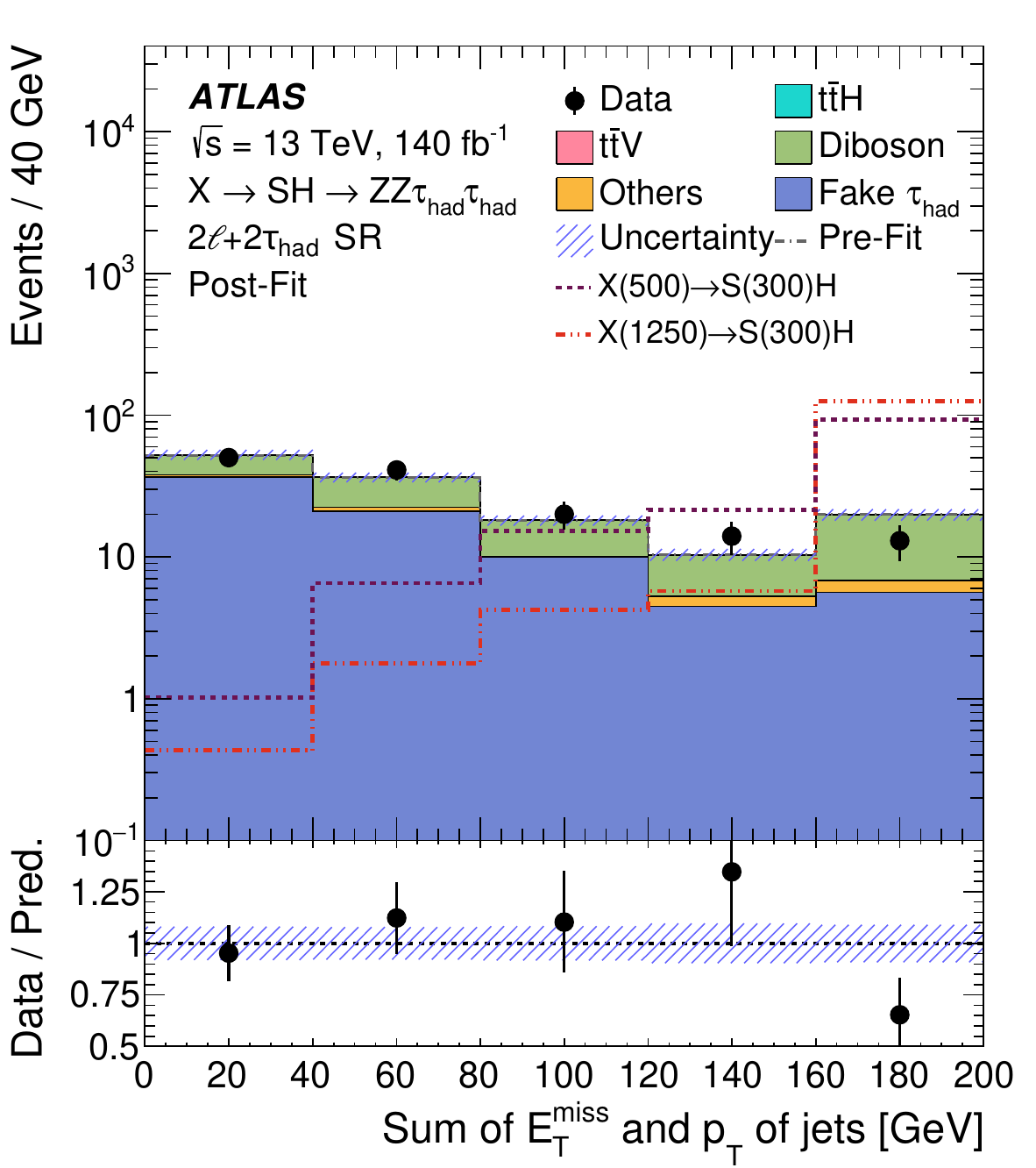} \\
(g) & (h) & (i) \\
\end{tabular}
\caption{Kinematic distributions obtained after a background-only fit to data (Post-fit) in the $\wwlditau$ signal region: $m_{\tau\tau}$ (a), leading \tauh $\pt$ (b), $\Delta R(\ell,\text{j})$ (c); in the $\wwllditau$ signal region: $m_{\tau\tau}$ (d), leading $\tauh$ $\pt$ (e), $\Delta R$($\ell,\ell$) (f); in the $\zzllditau$ signal region: $m_{\tau\tau}$ (g), leading \tauh $\pt$ (h), sum of $\met$ and $\pt$ of jets (i). The purple (red) dashed line represents the signal for $m_X=500$ \GeV\ and $m_S=300$ \GeV\ ($m_X=1250$ \GeV\ and $m_S=300$ \GeV) normalised to total background. The error band includes systematic and statistical uncertainties. The lower panel shows the ratio of data to post-fit background.}
\label{fig:plots_vv2taupostfit}
\end{figure}


\section{Systematic Uncertainties}
\label{sec:systematics}
Systematic uncertainties can affect both the normalisation of signal
and background and the shapes of their corresponding discriminant distributions.
Each source of uncertainty is considered to be uncorrelated from other sources.  For a given
source, the uncertainties across all processes and channels are treated as correlated.
 
\subsection{Luminosity}
The uncertainty in the integrated luminosity is 0.83\%, which affects the overall normalisation of all processes estimated from the simulation.
It is derived following a methodology similar to that detailed in Ref.~\cite{DAPR-2021-01X} 
and uses the LUCID-2 detector
for the baseline luminosity measurement \cite{Avoni:2018iuv} and a calibration of the luminosity scale from $x$--$y$ beam-separation scans.
 
\subsection{Reconstructed objects}
 
Uncertainties associated with electrons, muons, and $\tauh$ candidates arise from the trigger, reconstruction, momentum scale and resolution,
particle identification, and, in the case of electrons and muons, isolation efficiencies.
These are measured using $Z\to \ell^+\ell^-$ and $J/\psi\to \ell^+\ell^-$ events~\cite{PERF-2017-01,PERF-2015-10}
in the case of electrons and muons, and using $Z\to \tau^+\tau^-$ events in the case of $\tauh$ candidates~\cite{ATLAS-CONF-2017-029}.
 
Uncertainties associated with jets arise from the jet energy scale (JES)
and resolution (JER), and the efficiency to satisfy JVT requirements. 
The largest contribution comes from the jet energy scale, whose uncertainty dependence on jet $\pt$ and $\eta$, jet flavour,
and pile-up treatment is split into 43 uncorrelated components that are treated independently~\cite{PERF-2016-04}.
The total JES uncertainty is
below 5\% for most jets and below 1\% for central jets with $\pt$ between 300~\GeV\ and 2~\TeV. The difference between the JER values
in data and simulated events is treated as a systematic uncertainty with one nuisance parameter. It is applied to the simulated events by smearing the jet $\pt$ by the prescribed uncertainty.
 
Uncertainties associated with energy scales and resolutions of leptons and jets
are propagated to $\met$. Additional uncertainties originating from the modelling
of the underlying event, particularly its impact on the $\pt$ scale and resolution of unclustered energy, are negligible.
 
Efficiencies to veto tagged $b$-jets and $c$-jets in the simulation are corrected by $\pt$-dependent factors to match the efficiencies in data,
while the light-jet efficiency is scaled by $\pt$- and $\eta$-dependent factors.
The $b$-jet efficiency is measured in a data sample enriched in $\ttbar$ events~\cite{FTAG-2018-01}, while the $c$-jet efficiency is measured
using $\ttbar$ events~\cite{ATLAS-CONF-2018-001} or $W$+$c$-jet events~\cite{Aad:2015ydr}.
The light-jet efficiency is measured in a multijet data sample enriched in light-flavour jets~\cite{ATLAS-CONF-2018-006}.
These systematic uncertainties are taken to be uncorrelated between $b$-jets, $c$-jets, and light jets.
 
Pile-up reweighting scale factors are applied to simulated events to reproduce the pile-up distribution corresponding to the data.
The uncertainty in these scale factors is estimated by changing the profile in data within uncertainties.
 
\subsection{Background modelling}
\label{sec:syst_bkgmodeling}
 
Several sources of systematic uncertainty affecting the modelling of background are considered:
the choice of renormalisation and
factorisation scale in the matrix-element calculation, the choice of matching scale when matching the matrix elements to the parton
shower generator, and the uncertainty in the value of $\alpha_\text{s}$ when modelling initial-state radiation (ISR).
The uncertainty due to the choice of parton shower (PS) and hadronisation model is derived by comparing the predictions from \textsc{Powheg-Box} interfaced either to \textsc{Pythia\,8} or \textsc{Herwig\,7}. The latter uses the \textsc{MMHT2014 LO}~\cite{Harland-Lang:2014zoa} PDF set in combination with the H7UE tune~\cite{Bellm:2015jjp}. The uncertainty in the modelling of additional radiation from the PS is assessed by varying the corresponding parameter of the A14 set~\cite{ATL-PHYS-PUB-2016-004} and by varying the radiation renormalisation and factorisation scales independently by factors of 2.0 and 0.5. The PDF uncertainty is estimated by taking the PDF
variations as an envelope around the nominal PDF.
 
Uncertainties affecting the normalisation of the $V$+jets background are estimated
to be approximately 30\% by taking the maximum difference between the data and the total background prediction
in three or more jets~\cite{STDM-2018-49}.
This is mainly driven by different simulations used in the $V$+jets regions where \textsc{Sherpa} 2.2.1 at NLO is used for up to two partons, and LO is used for three or more partons.
 
Uncertainties in the $t\bar{t}$+jets background normalisation include those estimated from variations of the renormalisation and factorisation scales, the choice of parton shower and hadronisation, the \textsc{NNPDF3.0 NNLO} PDF set uncertainties, and $\alpha_\text{s}$ variations. The contribution of $t\bar{t}$+jets is small after the $b$-tagging veto is applied.
 
Uncertainties affecting the modelling of the single-top-quark background include a
$+5\%$/$-4\%$ uncertainty in the total cross section, which is estimated as a weighted average
of the theoretical uncertainties in $t$-, $tW$- and $s$-channel production~\cite{Kidonakis:2011wy,Kidonakis:2010ux,Kidonakis:2010tc}.
Additional uncertainties associated with the parton shower, hadronisation, and initial- and final-state radiation are also considered.
 
The diboson normalisation uncertainty is estimated to be $\pm 15\%$, which includes those estimated from variations of the renormalisation and factorisation scales, \textsc{NNPDF3.0 NNLO} PDF set, and $\alpha_\text{s}$ variations. Uncertainties in the $\ttbar V$, $\ttbar H$, and VH cross sections are estimated to be $\pm$10\%, [$+5.8\%/-9.2\%$], and $\pm$1.4\%, respectively, arising from the uncertainties in their respective NLO theoretical cross sections~\cite{TOPQ-2018-08}.
Most of the rare background contributions ($tZ$, $t\bar{t}t$, $t\bar{t}WW$, $WtZ$, $VVV$) are normalised using their NLO theoretical cross sections,
and assigned a 50\% normalisation uncertainty, except for $tZ$ where a 5\% normalisation uncertainty is used.
 
The statistical uncertainties of the fake \tauh background calibration are applied with uncorrelated uncertainties for different sources of fake $\tau$-leptons. The uncertainties in the modelling of the processes used to evaluate the fake factors are taken into account in the fit.
 
\subsection{Signal modelling}
Several normalisation and shape uncertainties are considered and they are treated as fully correlated for the $X\to SH$ signals.
Uncertainties in the $S$ and $H$ decay branching ratios are considered by following the recommendation in Ref.~\cite{deFlorian:2016spz}.
The BDT variations due to ISR, FSR, scale, and PDF uncertainties are estimated from a non-resonant $HH$ sample and
are applied to the $SH$ signal as fully correlated in all SRs.
The PS uncertainty is also estimated by comparing the
BDT shape from a $HH$ sample with an alternative sample interfaced with \textsc{Herwig\,7} and is applied to the $SH$ signal.
This choice is motivated by the fact that the modelling of the non-resonant $HH$ signal is calculated to higher order
than the $SH$ simulation, and their kinematics are quite similar to those in the low $m_X$ mass points.
These uncertainties are cross-checked and found to be consistent with those calculated in the resonant $X\to HH\to bb\tau\tau$ analysis~\cite{ATLAS:2022xzm}.


\section{Statistical analysis}
\label{sec:stat}
The final discriminant distributions are the BDT outputs from all considered analysis regions. They are jointly analysed for the
presence of a signal. The BDT distributions are binned to maximise the sensitivity to the signal and the binning is optimised to reduce the statistical fluctuations.
 
The statistical analysis uses a binned likelihood function ${\cal L}(\mu,\theta)$ constructed as
a product of Poisson probability terms over all bins considered in the search\cite{Cowan:2010js}. This function depends
on the signal-strength parameter $\mu$, defined as a factor multiplying the expected yield of signal events
normalised to an arbitrary reference cross section $\sigma(pp\to X\to SH)=$~1~pb,
and on $\theta$, a set of nuisance parameters (NPs) that encode the effect of systematic uncertainties in
the signal and background expectations. The expected total number of events in a given bin depends on $\mu$ and $\theta$.
All nuisance parameters are subject to Gaussian constraints in the likelihood function.
For a given value of $\mu$, the NPs $\theta$ allow variations of the expectations for signal and background
consistent with the corresponding systematic uncertainties, hence their fitted values result in deviations from
the nominal expectations that globally provide the best fit to the data.
This procedure reduces the impact of systematic uncertainties on
the sensitivity of the search by taking advantage of the highly populated background-dominated bins included in the likelihood fit.
Statistical uncertainties in each bin of the predicted final discriminant distributions are considered through
dedicated parameters (gammas) in the fit.
The best-fit $\mu$ is obtained by performing a binned likelihood fit to the data under the signal-plus-background
hypothesis and maximising the likelihood function ${\cal L}(\mu,\theta)$ over $\mu$ and $\theta$.
 
The fitting procedure was initially validated through extensive studies using pseudo-data (which is defined as the sum of all predicted backgrounds plus an injected signal of variable strength) and by performing fits to the data considering only those bins of the final discriminating variable that lie in the low BDT signal-depleted regions defined by BDT~$<$~0.1. In both cases, the robustness of the model for systematic uncertainties was established by verifying the stability of the fitted background when varying the leading sources of uncertainty. After this, the whole BDT distribution in the data is fitted under the signal-plus-background
hypothesis. Further checks involve the comparison of the fitted nuisance parameters before and after removal of the blinding requirements. In addition, it is verified that the fit is able to determine the strength of a simulated signal injected into the real data.
 
The test statistic $q_\mu$ is defined as the profile likelihood ratio,
$q_\mu = -2\ln({\cal L}(\mu,{\hat{\hat{\theta}}}_\mu)/{\cal L}(\hat{\mu},\hat{\theta}))$,
where $\hat{\mu}$ and $\hat{\theta}$ are the values of the parameters that
maximise the likelihood function (subject to the constraint $\hat{\mu} \geq 0$), and ${\hat{\hat{\theta}}}_\mu$ are the values of NPs that maximise the likelihood function for a given value of $\mu$. The test statistic $q_\mu$ is evaluated with the {\textsc RooFit} package~\cite{RooFit}.
 
Exclusion limits are set on $\mu$, derived by using $q_\mu$ in the CL$_{\textrm{s}}$ method~\cite{Junk:1999kv,Read:2002hq}.
For a given signal scenario, values of $\mu$ yielding CL$_{\textrm{s}} < 0.05$,
where CL$_{\textrm{s}}$ is computed using the asymptotic approximation~\cite{Cowan:2010js}, are excluded with at least 95\% confidence.
Exclusion limits are also verified using pseudo-experiments, which give consistent results to within 25\%.


\section{Results}
\label{sec:results}
 
A binned likelihood fit under the signal-plus-background hypothesis is performed on each BDT discriminant distribution
for each mass point ($m_X$, $m_S$), in the three signal regions, in the two $WW$ regions combined, and in all three regions combined.
This gives a total of $5\times 18=90$ fits. The unconstrained parameter of the fit is the signal strength ($\mu$).
 
Figure~\ref{fig:bdt_postfit} shows the BDT output distributions after the background-only fit to the
data for the $X\to SH$ searches for the most sensitive high mass point ($m_X=1250$ \GeV, $m_S=300$ \GeV) and
the least sensitive low mass point ($m_X=500$ \GeV, $m_S=300$ \GeV), respectively.
No significant pulls or constraints are obtained for the fitted nuisance parameters, resulting in a post-fit background
prediction in each signal region that is very close to the pre-fit prediction, except with reduced uncertainties resulting from the fit.
 
No excess of events is observed beyond the expected SM background, hence 95\% CL upper limits are set on the production cross section
$\sigma(pp\to X\to SH)$ as shown in Table~\ref{tab:xseclimits_combined} for the combined $WW$ regions, the $ZZ$ region, and all regions combined after correcting the $S\to WW,\, ZZ$ branching fractions assuming the $S$ decays like a SM Higgs boson~\cite{cyrm-2017-002}.
The combined limit is dominated by the $\wwlditau$ channel and is improved by 26\%--53\% after adding the $\wwllditau$
and $\zzllditau$ channels.
The results are dominated by the statistical uncertainty. The main contributions to the total systematic uncertainty arise from the tau-lepton fakes, the diboson modelling, the $\tauh$ identification efficiency, lepton identification,
and trigger efficiencies. Their impact on the cross section $\sigma(pp\to X\to SH)$ is summarized in Table~\ref{tab:sys_impact} for the least- and most-sensitive mass points.
The impact on the combined limits is between 2\% and 14\% across all mass points.
 
The best expected combined limit is 85 fb for $\sigma(pp\to X\to SH)$ for $m_X=1250$ \GeV\ and $m_S=300$ \GeV.
The observed and expected 95\% CL upper limits are summarized in Figure~\ref{fig:xseclimits_masses} for $\sigma(pp\to X\to SH)$, $\sigma(pp\to X\to SH\to WW\tau^+\tau^-)$,
and $\sigma(pp\to X\to SH\to ZZ\tau^+\tau^-)$ as a function of combined $m_S$ and $m_X$ masses ($m_S$+$m_X/25$) in \GeV.
The maximum values allowed for $\sigma(pp\to X\to SH\to WW\tau^+\tau^-)$ and $\sigma(pp\to X\to SH\to ZZ\tau^+\tau^-)$ as a function of $m_X$ and $m_S$ in the NMSSM parameter space are also shown for comparison. These values
are obtained via a parameter space scan using \textsc{NMSSMTOOLS 6.0.0}~\cite{Ellwanger_2005,Ellwanger_2006,Allanach_2009} and \textsc{NMSSMCALC}~\cite{Baglio_2014}, following recommendations from Ref.~\cite{deFlorian:2016spz}.
The scan uses measurements of Higgs boson properties, searches for supersymmetry, B-meson physics, and searches for dark matter to exclude cross section values above the indicated line.
The observed limits approach the allowed cross sections in the low-$m_X$ and low-$m_S$ part of the NMSSM parameter space. This NMSSM scan for $VV\tau\tau$ is also compatible with other NMSSN scans
using  $X\to SH\to bbbb$, $bb\tau\tau$, and $bb\gamma\gamma$ final states.
 
The corresponding 2D limits are also shown in Figure~\ref{fig:xseclimits_masses2d} as a function of $m_X$ and $m_S$ masses in \GeV.
The limit at each mass point is compared to the limits obtained using the BDTs trained for the (lower or upper) neighbouring mass values
in the fit and found to be consistent within 15\%, indicating that interpolating the limits between mass points is unnecessary.
The fit results are also used to set limits on the cross sections for $X\to SH\to VV\tau\tau$ decays in events
with a pair of $\tau$-leptons and multiple light-leptons.

\begin{figure}[htbp]
\centering
\begin{tabular}{@{}ccc@{}}
\includegraphics[width=0.28\textwidth]{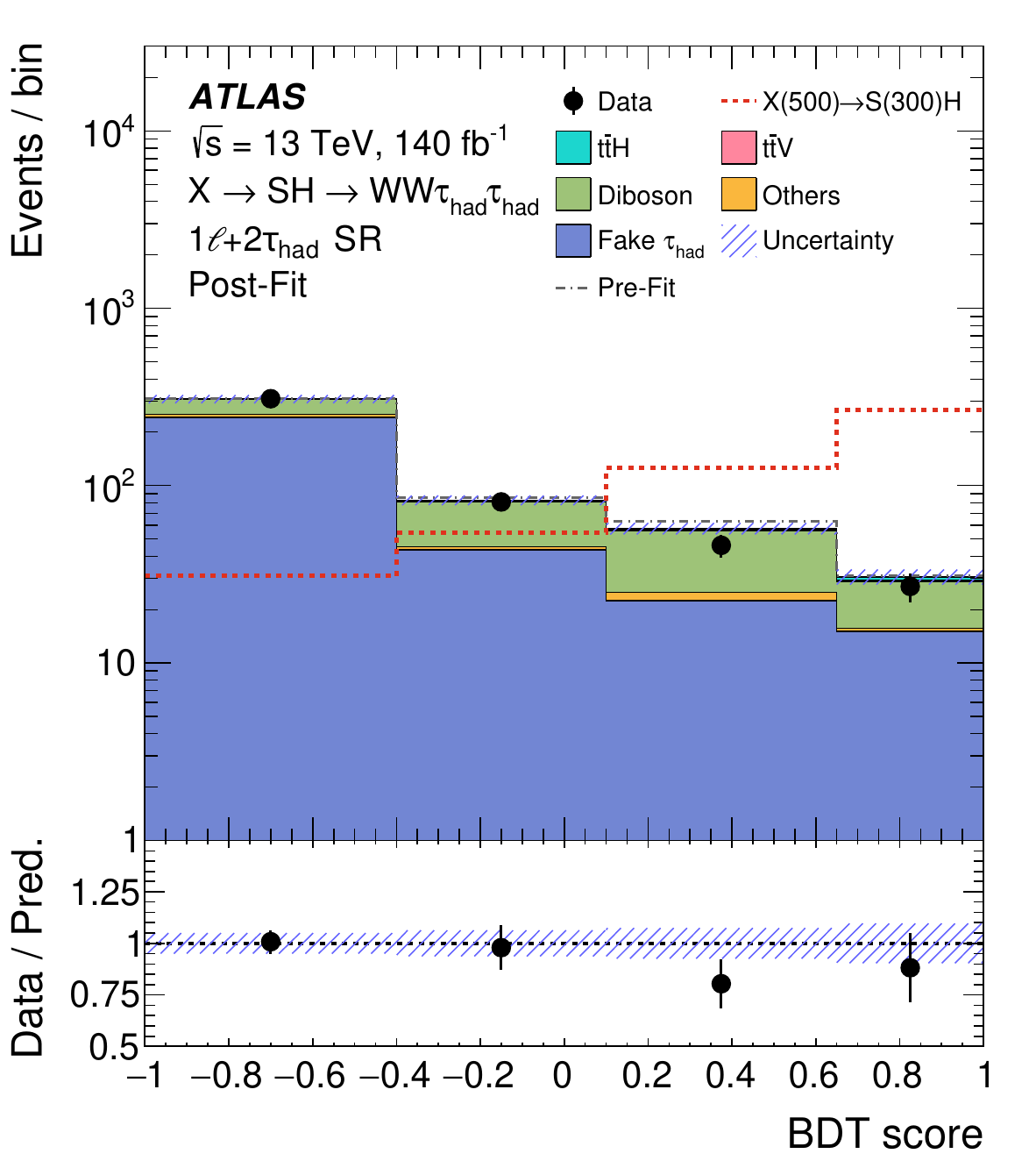} & 
\includegraphics[width=0.28\textwidth]{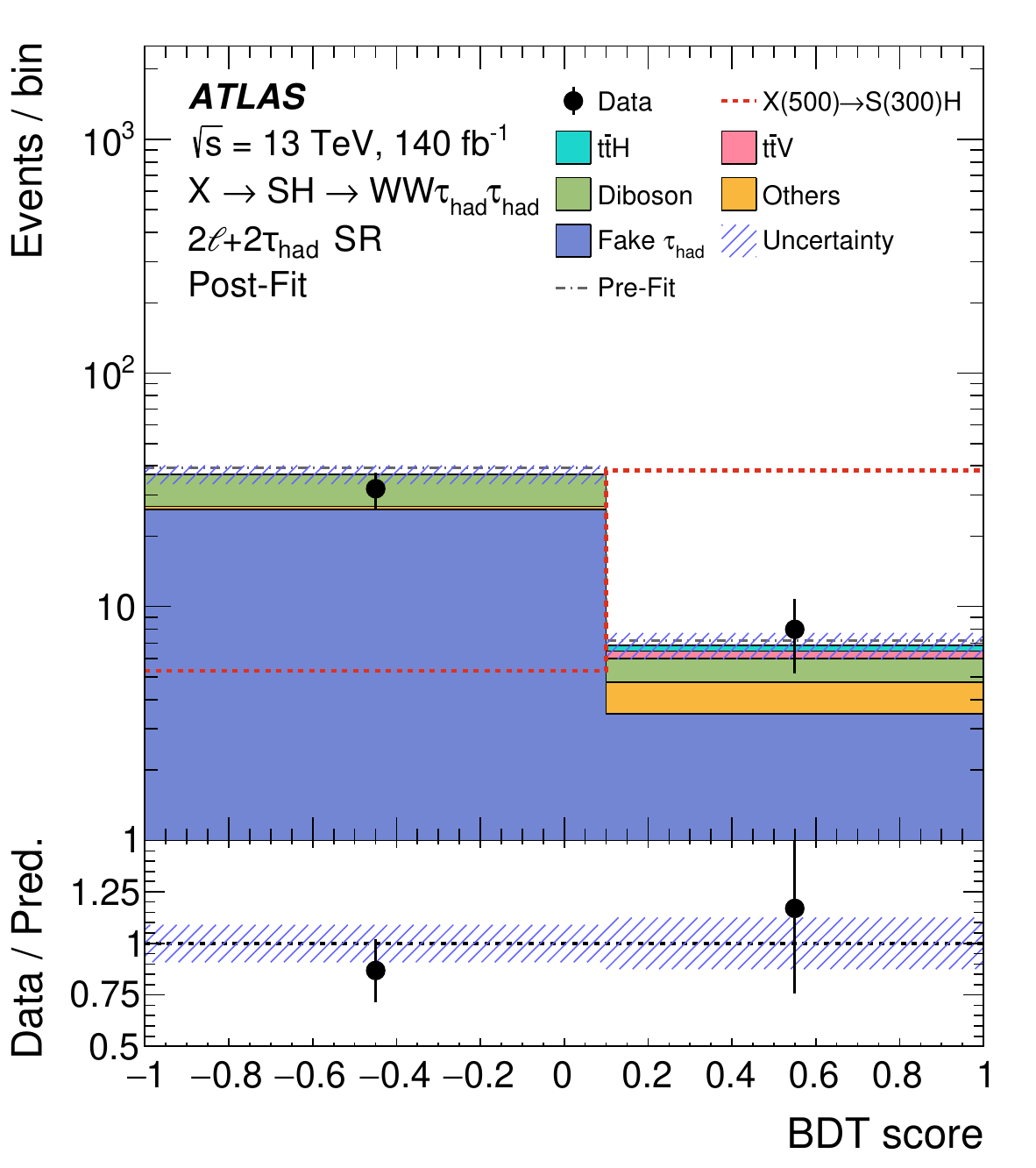} & 
\includegraphics[width=0.28\textwidth]{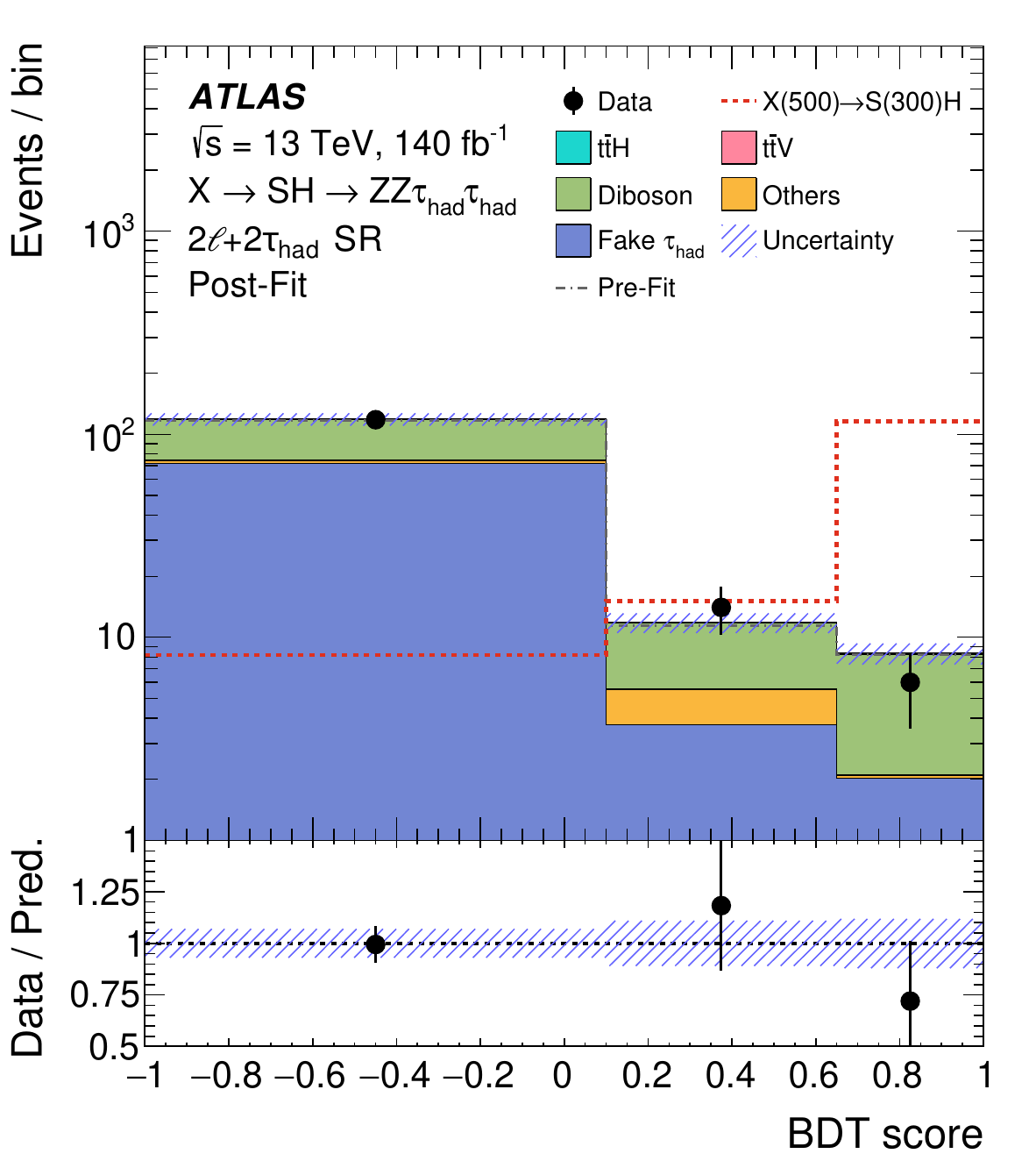} \\ 
(a) & (b) & (c) \\
\end{tabular}
\begin{tabular}{@{}ccc@{}}
\includegraphics[width=0.28\textwidth]{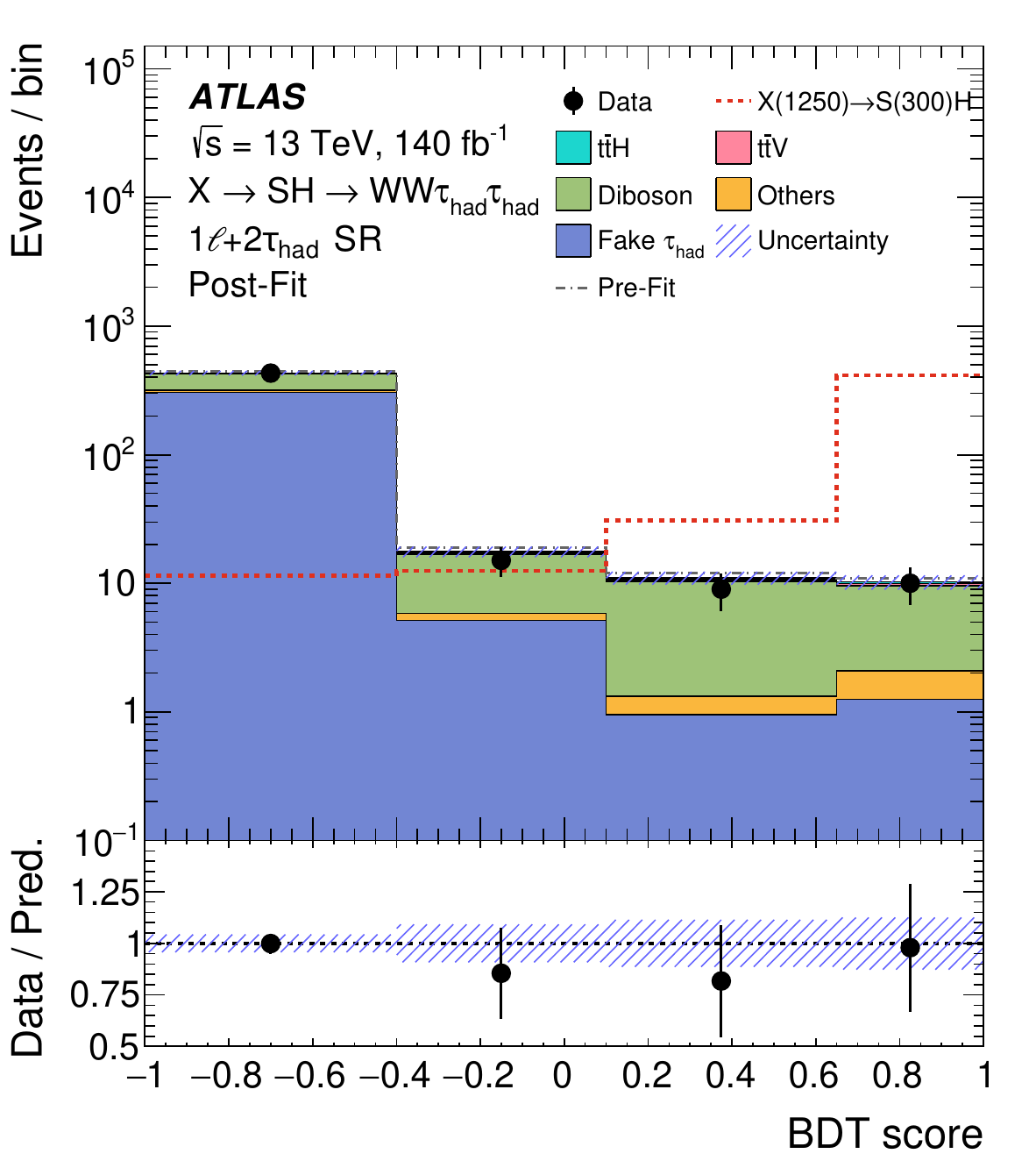} &
\includegraphics[width=0.28\textwidth]{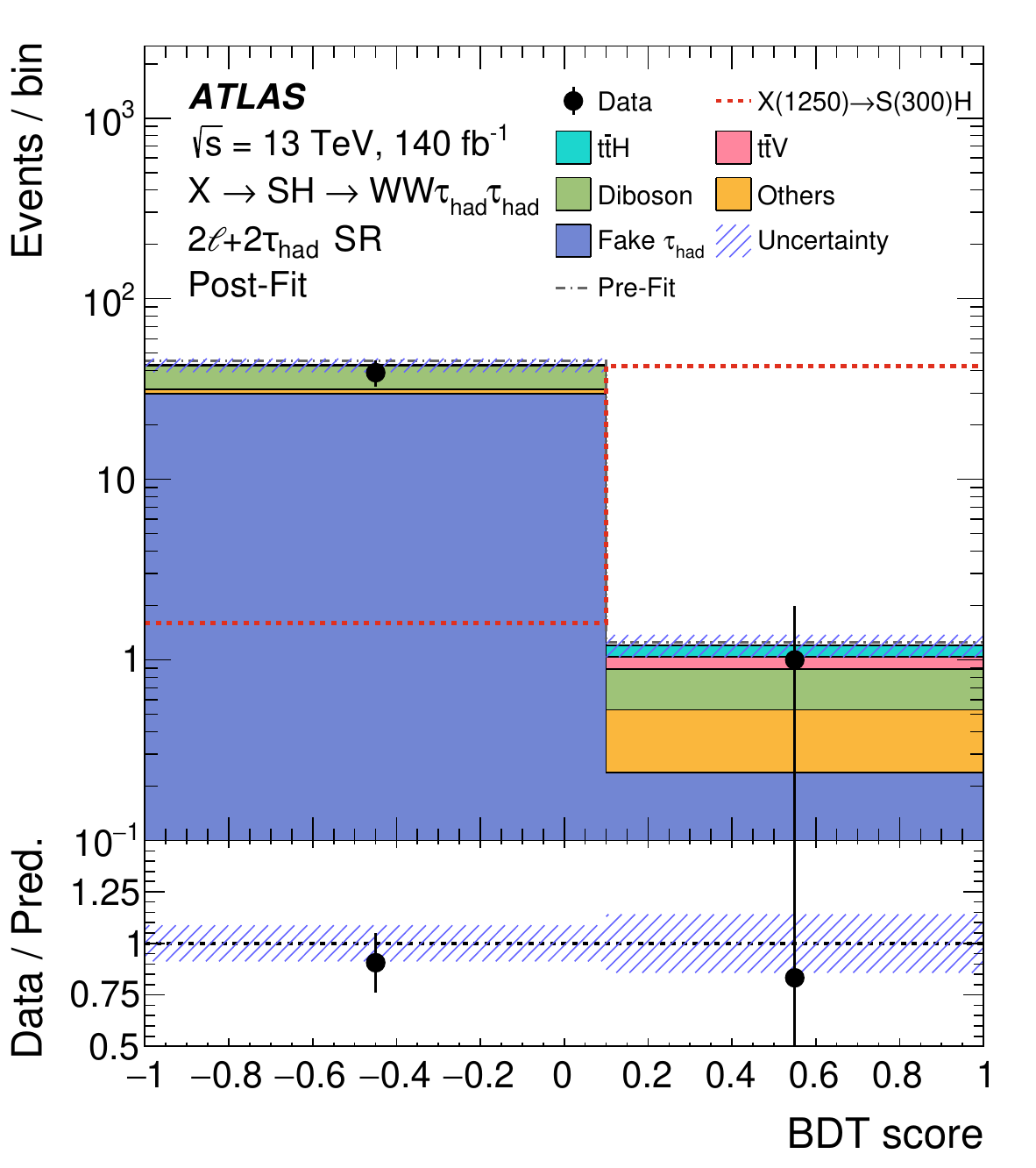} & 
\includegraphics[width=0.28\textwidth]{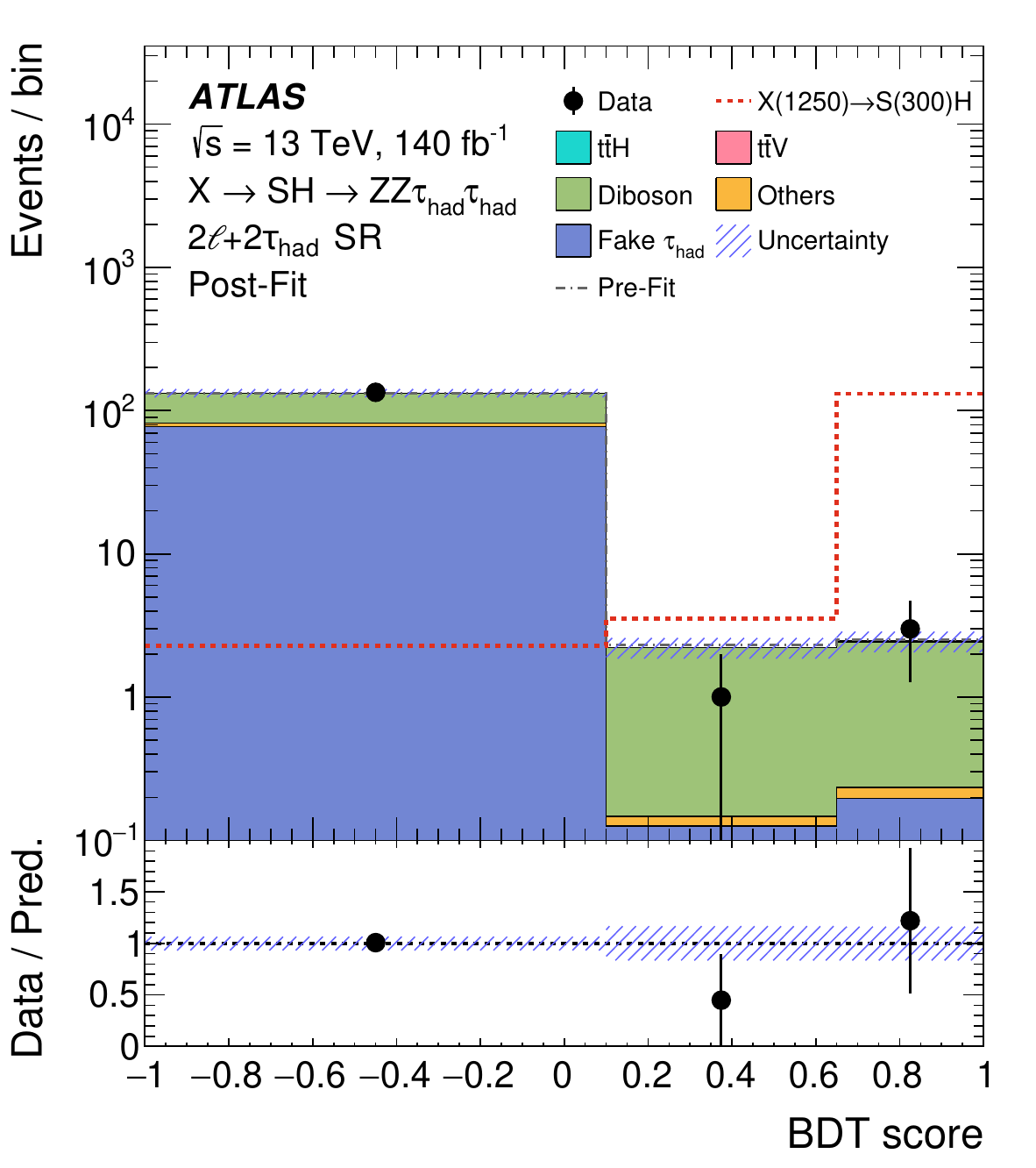} \\
(d) & (e) & (f)\\
\end{tabular}
\caption{BDT output distributions obtained from a background-only fit to the data for the $X\to SH$ search with $m_X=500$ \GeV\ and $m_S=300$ \GeV\ in the channel of (a) $\wwlditau$, (b) $\wwllditau$, (c) $\zzllditau$; and with $m_X=1250$ \GeV\ and $m_S=300$ \GeV\ in the channel of (d) $\wwlditau$, (e) $\wwllditau$, (f) $\zzllditau$. The red dashed line represents the signal normalised to total background. The error band includes systematic and statistical uncertainties.}
\label{fig:bdt_postfit}
\end{figure}
 
\begin{figure}[htbp]
\centering
\subfloat[]{\includegraphics[width=0.7\textwidth]{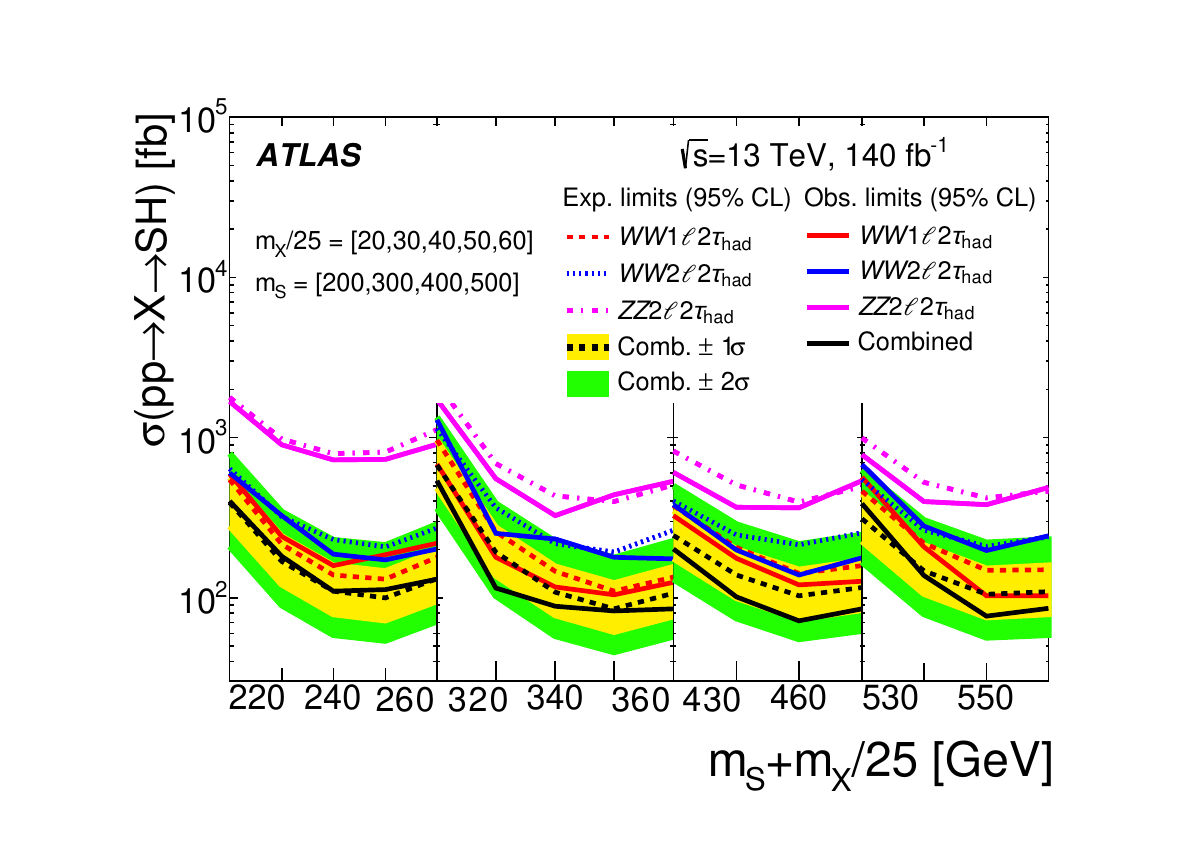}} \\
\subfloat[]{\includegraphics[width=0.49\textwidth]{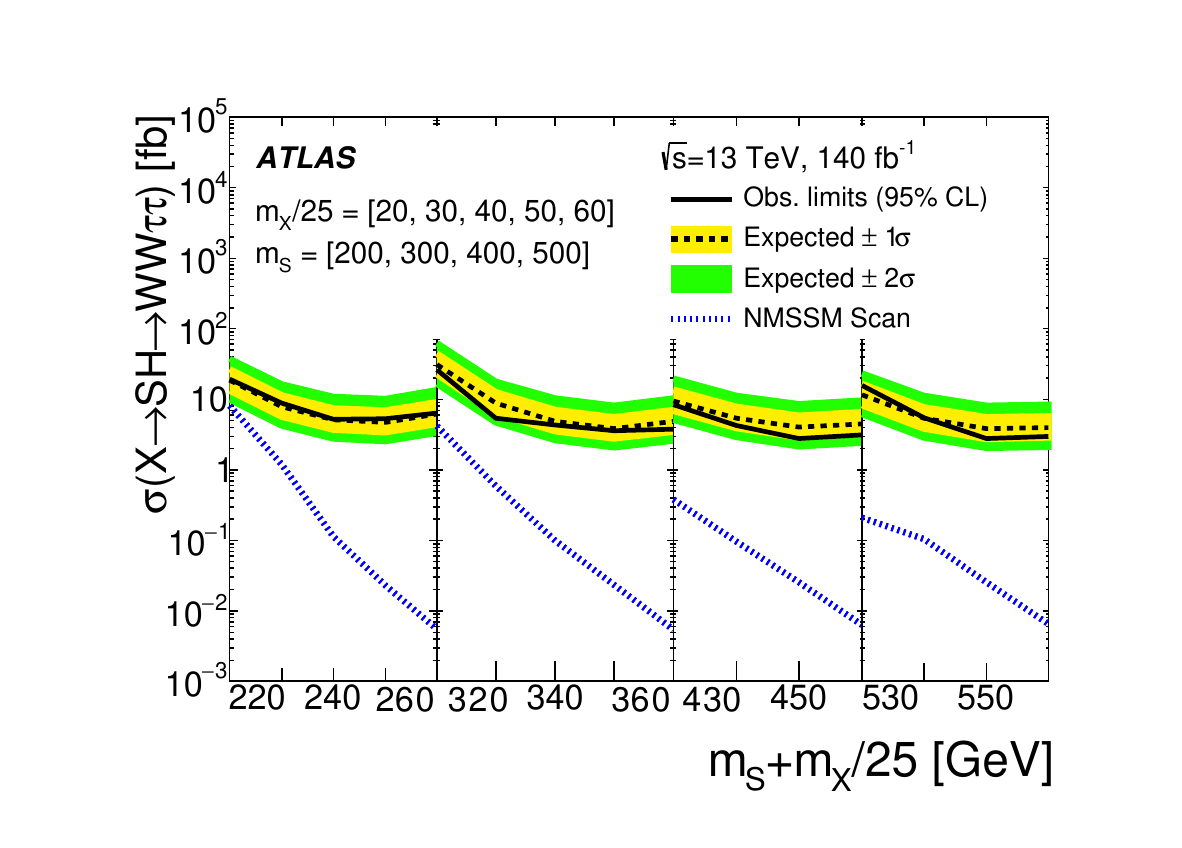}}
\subfloat[]{\includegraphics[width=0.49\textwidth]{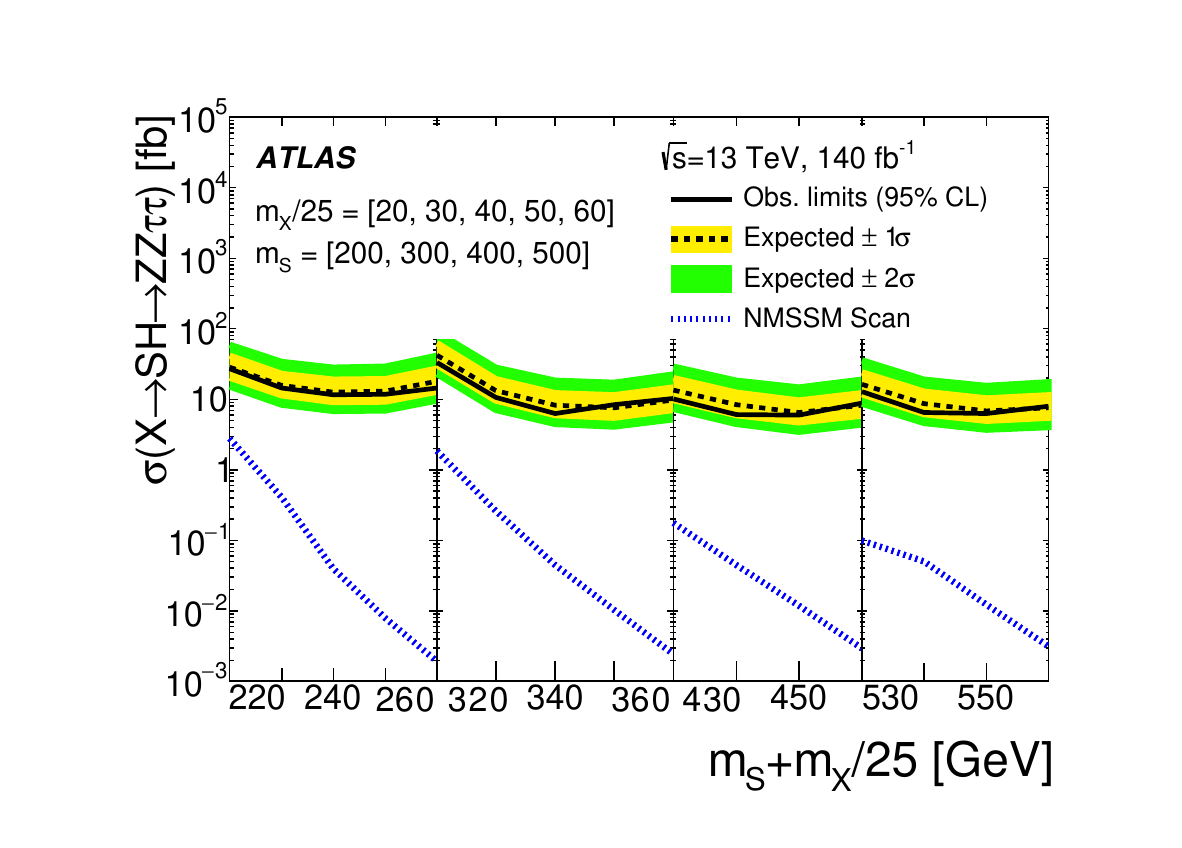}}
\caption{Observed and expected 95\% CL upper limits are shown for (a) $\sigma(pp\to X\to SH)$ obtained from three channels and their combination; (b) $\sigma(pp\to X\to SH\to WW\tau^+\tau^-)$ obtained from the combination of $\wwlditau$ and $\wwllditau$ channels; (c) $\sigma(pp\to X\to SH\to ZZ\tau^+\tau^-)$ obtained from $\zzllditau$ channel, as a function of combined $m_S$ and $m_X$ masses ($m_S$+$m_X/25$) in \GeV. The NMSSM scans of the allowed cross sections for $\sigma(pp\to X\to SH\to WW\tau^+\tau^-)$ and $\sigma(pp\to X\to SH\to ZZ\tau^+\tau^-)$ are also shown.}
\label{fig:xseclimits_masses}
\end{figure}
 
\begin{figure}[htbp]
\centering
\subfloat[]{\includegraphics[width=0.7\textwidth]{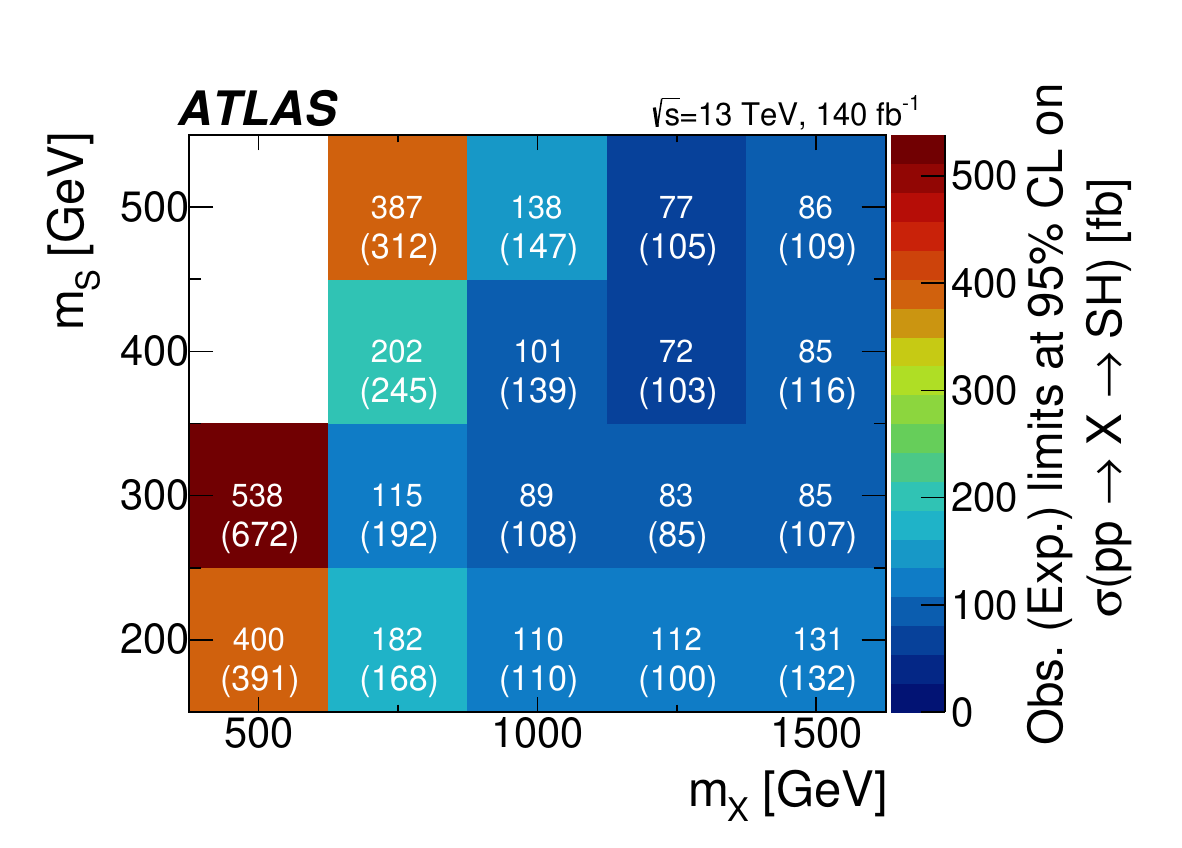}} \\
\subfloat[]{\includegraphics[width=0.49\textwidth]{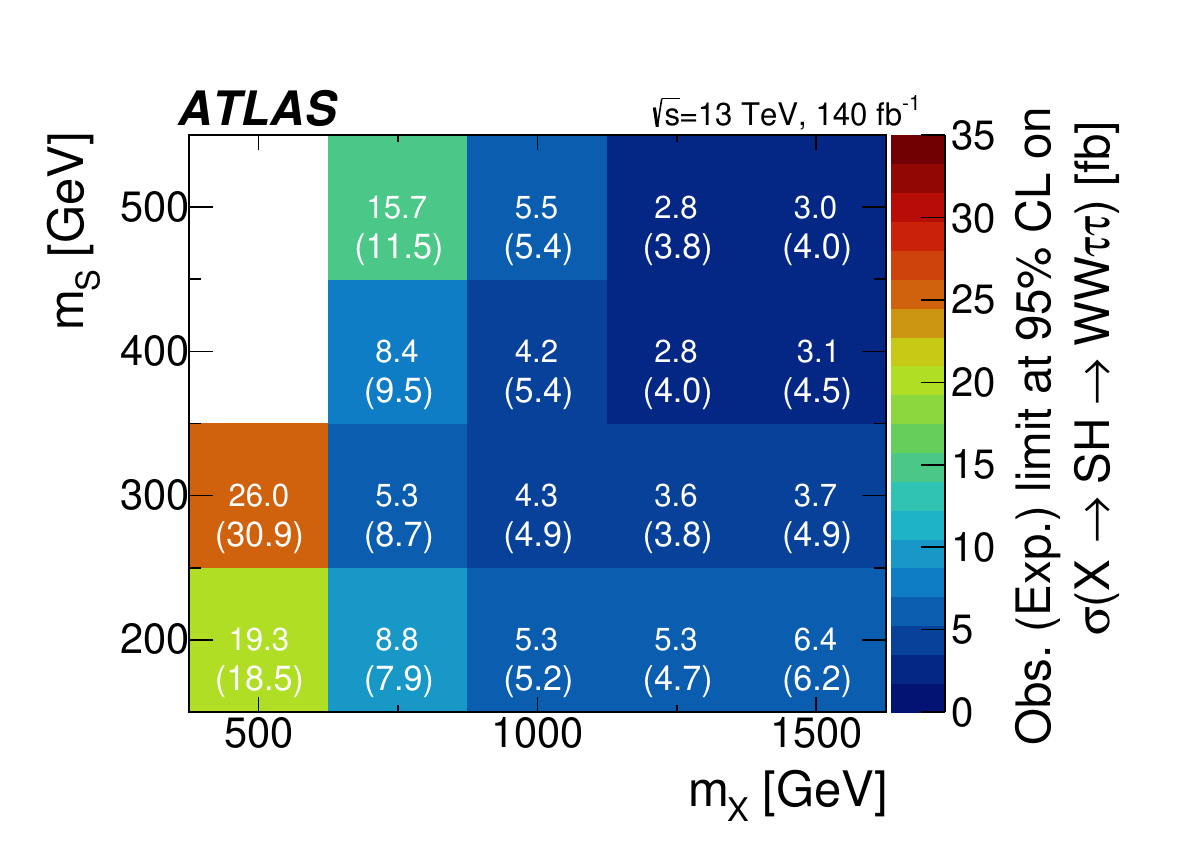}}
\subfloat[]{\includegraphics[width=0.49\textwidth]{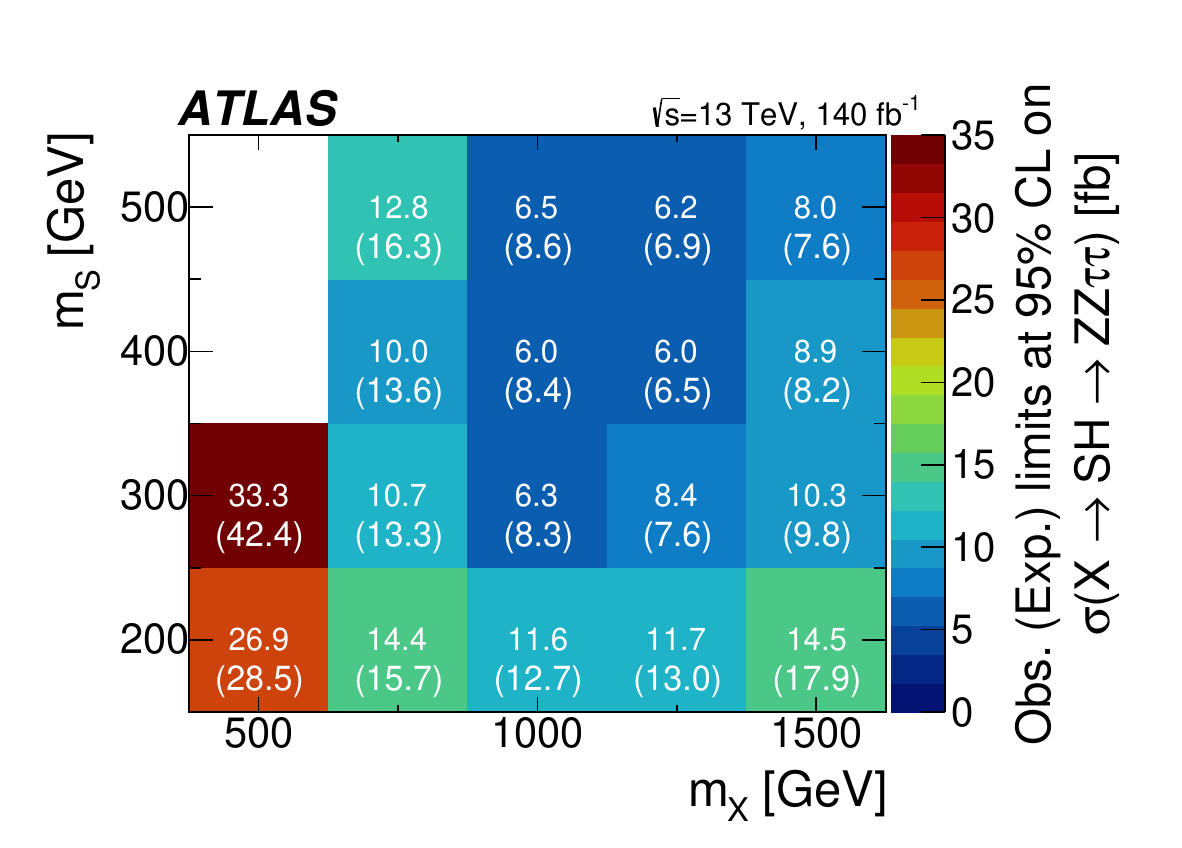}}
\caption{Observed (expected) 2D limits are shown for (a) $\sigma(pp\to X\to SH)$, (b) $\sigma(pp\to X\to SH\to WW\tau^+\tau^-)$, and (c) $\sigma(pp\to X\to SH\to ZZ\tau^+\tau^-)$ as a function of $m_X$ and $m_S$ masses in \GeV.}
\label{fig:xseclimits_masses2d}
\end{figure}
 
\begin{table}[htbp]
\caption{Summary of observed and expected 95\% CL upper limits on the production cross section of $\sigma(pp\to X\to SH)$ in fb in the combination of $WW+ZZ$, and in the $WW$ and $ZZ$ individual channels. The best expected limit is
85~fb for $m_X=1250$ \GeV\ and $m_S =300$ \GeV. The combination limits gain about 26\%--53\% over the best individual limit from the $\wwlditau$ channel.}
\label{tab:xseclimits_combined}
\begin{center}
\resizebox{\textwidth}{!}{
\begin{tabular}{lc|c|c|c|c|c|c}\toprule\toprule
$m_X$ & $m_S$ &\multicolumn{2}{c|}{Combined $\sigma(pp\to X\to SH)$ [fb]} & \multicolumn{2}{c|}{$\sigma(pp\to X\to WW\tau\tau)$ [fb]} & \multicolumn{2}{c}{$\sigma(pp\to X\to ZZ\tau\tau)$ [fb]} \\ \midrule
(\GeV)& (\GeV) & Observed & Expected & Observed & Expected & Observed & Expected \\ \midrule
500 & 200 & 400 & $391^{+170}_{-110}$ & 19 & $18^{+8}_{-5}$ & 27 & $28^{+13}_{-8}$ \\
750 & 200 & 182 & $168^{+74}_{-47}$ & 8.8 & $8^{+3.5}_{-2}$ & 14 & $16^{+7}_{-4}$ \\
1000 & 200 & 110 & $110^{+49}_{-31}$ & 5.3 & $5^{+2}_{-1}$ & 12 & $13^{+6}_{-4}$ \\
1250 & 200 & 112 & $100^{+46}_{-28}$ & 5.4 & $5^{+2}_{-1}$ & 12 & $13^{+6}_{-4}$ \\
1500 & 200 & 131 & $132^{+62}_{-37}$ & 6.4 & $6^{+3}_{-2}$ & 14 & $18^{+9}_{-5}$ \\
\midrule
500 & 300 & 538 & $672^{+290}_{-190}$ & 26 & $31^{+13}_{-9}$ & 33 & $42^{+19}_{-12}$ \\
750 & 300 & 115 & $192^{+83}_{-54}$ & 5.3 & $9^{+4}_{-2}$ & 11 & $13^{+6}_{-4}$ \\
1000 & 300 & 88.6 & $108^{+48}_{-30}$ & 4.3 & $5^{+2}_{-1}$ & 6.3 & $8^{+4}_{-2}$ \\
1250 & 300 & 82.6 & $85^{+38}_{-24}$ & 3.6 & $4^{+2}_{-1}$ & 8.4 & $8^{+4}_{-2}$ \\
1500 & 300 & 85.4 & $107^{+49}_{-30}$ & 3.7 & $5^{+2}_{-1}$ & 10 & $10^{+5}_{-3}$ \\
\midrule
750 & 400 & 202 & $245^{+107}_{-68}$ & 8.4 & $10^{+4}_{-3}$ & 10 & $14^{+6}_{-4}$ \\
1000 & 400 & 101 & $139^{+62}_{-39}$ & 4.2 & $5^{+2}_{-1.5}$ & 6 & $8^{+4}_{-2}$ \\
1250 & 400 & 71.7 & $103^{+46}_{-29}$ & 2.8 & $4^{+2}_{-1}$ & 6 & $6.5^{+3}_{-2}$ \\
1500 & 400 & 85.1 & $116^{+53}_{-32}$ & 3.1 & $4.5^{+2}_{-1}$ & 8.9 & $8^{+4}_{-2}$ \\
\midrule
750 & 500 & 387 & $312^{+135}_{-87}$ & 16 & $11^{+5}_{-3}$ & 13 & $16^{+8}_{-4.5}$ \\
1000 & 500 & 138 & $147^{+65}_{-41}$ & 5.5 & $5^{+2}_{-1.5}$ & 6.5 & $9^{+4}_{-2}$ \\
1250 & 500 & 77 & $105^{+47}_{-29}$ & 2.8 & $4^{+2}_{-1}$ & 6.2 & $7^{+3}_{-2}$ \\
1500 & 500 & 85.9 & $109^{+50}_{-31}$ & 3 & $4^{+2}_{-1}$ & 8 & $8^{+4}_{-2}$ \\
\bottomrule\bottomrule
\end{tabular}
}
\end{center}
\end{table}

\begin{table}[h!]
\caption{Relative uncertainties in the observed 95\% CL upper limits of $\sigma(pp\to X\to SH)$ obtained from the combined fit to the data for the least- and most-sensitive mass points. The uncertainties are
symmetrized and grouped into the categories described in the systematic section.}
\label{tab:sys_impact}
\begin{center}
\begin{tabular}{
@{}l
S
S
@{}
}
\toprule\toprule
\multirow{2}{*}{Source of uncertainty}      & \multicolumn{2}{c}{$\Delta\sigma/\sigma(pp\to X \to SH)$ [\%]} \\
& \multicolumn{1}{c}{$m_X$ = 500, $m_S$ = 300 [$\GeV$]} & \multicolumn{1}{c}{$m_X$ = 1250, $m_S$ = 300 [$\GeV$]} \\\midrule
Lepton ID                               & 5           &1         \\
JES and JER                             & 6           &4         \\
MC modelling                             & 11           &9         \\
Fake $\tau$ modelling                    & 10           &5         \\
$\tau$ ID                               & 11           &6         \\
Luminosity and triggers                 & 3           &2         \\
MC (CR) statistics                      & 8           &5         \\\midrule
Total systematic uncertainty            & 27          &15        \\
Data statistical uncertainty            & 46          &40        \\\midrule
Total uncertainties                     & 53          &43          \\
\bottomrule\bottomrule
\end{tabular}
\end{center}
\end{table}

\clearpage


\section{Conclusion}
\label{sec:conclusion}
A search for a new heavy scalar particle $X$ decaying into a Standard Model (SM) Higgs boson (with $H\to \tau^+\tau^-$) and a
new singlet scalar particle $S$ ($S\to VV$ with $V=W,\,Z$) is presented. The search is based on a data sample of $p p$
collisions at $\sqs = 13$ \TeV\ corresponding to an integrated luminosity of 140 $\ensuremath{\textrm{fb}^{-1}}$ recorded with
the ATLAS detector at the Large Hadron Collider. The explored $X$ masses range from 500 to 1500 \GeV, with
the corresponding $S$ mass in the range 200--500 \GeV. The search selects events with two hadronically
decaying $\tau$-lepton candidates from $H\to \tau^+\tau^-$ decays and one or two light leptons
($\ell=e,\,\mu$) from $S\to VV$ decays. A multivariate discriminant based on event kinematics is used to separate the signal from the background.
No excess of events is observed beyond the expected SM background
and 95\% CL upper limits on the cross-section for $\sigma(pp\to X\to SH)$, assuming the same SM-Higgs boson-like decay branching ratios
for $S\to VV$ decay, are derived between
72 fb and 542 fb.
Upper limits on the visible cross-sections $\sigma(pp\to X\to SH \to WW\tau\tau)$
and $\sigma(pp\to X\to SH \to ZZ\tau\tau)$ are set
in the ranges 3--26 fb and 5--33 fb, respectively.


\section*{Acknowledgements}


We thank CERN for the very successful operation of the LHC, as well as the
support staff from our institutions without whom ATLAS could not be
operated efficiently.
 
We acknowledge the support of
ANPCyT, Argentina;
YerPhI, Armenia;
ARC, Australia;
BMWFW and FWF, Austria;
ANAS, Azerbaijan;
CNPq and FAPESP, Brazil;
NSERC, NRC and CFI, Canada;
CERN;
ANID, Chile;
CAS, MOST and NSFC, China;
Minciencias, Colombia;
MEYS CR, Czech Republic;
DNRF and DNSRC, Denmark;
IN2P3-CNRS and CEA-DRF/IRFU, France;
SRNSFG, Georgia;
BMBF, HGF and MPG, Germany;
GSRI, Greece;
RGC and Hong Kong SAR, China;
ISF and Benoziyo Center, Israel;
INFN, Italy;
MEXT and JSPS, Japan;
CNRST, Morocco;
NWO, Netherlands;
RCN, Norway;
MEiN, Poland;
FCT, Portugal;
MNE/IFA, Romania;
MESTD, Serbia;
MSSR, Slovakia;
ARRS and MIZ\v{S}, Slovenia;
DSI/NRF, South Africa;
MICINN, Spain;
SRC and Wallenberg Foundation, Sweden;
SERI, SNSF and Cantons of Bern and Geneva, Switzerland;
MOST, Taiwan;
TENMAK, T\"urkiye;
STFC, United Kingdom;
DOE and NSF, United States of America.
In addition, individual groups and members have received support from
BCKDF, CANARIE, Compute Canada and CRC, Canada;
PRIMUS 21/SCI/017 and UNCE SCI/013, Czech Republic;
COST, ERC, ERDF, Horizon 2020 and Marie Sk{\l}odowska-Curie Actions, European Union;
Investissements d'Avenir Labex, Investissements d'Avenir Idex and ANR, France;
DFG and AvH Foundation, Germany;
Herakleitos, Thales and Aristeia programmes co-financed by EU-ESF and the Greek NSRF, Greece;
BSF-NSF and MINERVA, Israel;
Norwegian Financial Mechanism 2014-2021, Norway;
NCN and NAWA, Poland;
La Caixa Banking Foundation, CERCA Programme Generalitat de Catalunya and PROMETEO and GenT Programmes Generalitat Valenciana, Spain;
G\"{o}ran Gustafssons Stiftelse, Sweden;
The Royal Society and Leverhulme Trust, United Kingdom.
 
The crucial computing support from all WLCG partners is acknowledged gratefully, in particular from CERN, the ATLAS Tier-1 facilities at TRIUMF (Canada), NDGF (Denmark, Norway, Sweden), CC-IN2P3 (France), KIT/GridKA (Germany), INFN-CNAF (Italy), NL-T1 (Netherlands), PIC (Spain), ASGC (Taiwan), RAL (UK) and BNL (USA), the Tier-2 facilities worldwide and large non-WLCG resource providers. Major contributors of computing resources are listed in Ref.~\cite{ATL-SOFT-PUB-2021-003}.


\clearpage
 
\clearpage
\printbibliography
 
\clearpage
 
\begin{flushleft}
\hypersetup{urlcolor=black}
{\Large The ATLAS Collaboration}

\bigskip

\AtlasOrcid[0000-0002-6665-4934]{G.~Aad}$^\textrm{\scriptsize 102}$,
\AtlasOrcid[0000-0002-5888-2734]{B.~Abbott}$^\textrm{\scriptsize 120}$,
\AtlasOrcid[0000-0002-1002-1652]{K.~Abeling}$^\textrm{\scriptsize 55}$,
\AtlasOrcid[0000-0001-5763-2760]{N.J.~Abicht}$^\textrm{\scriptsize 49}$,
\AtlasOrcid[0000-0002-8496-9294]{S.H.~Abidi}$^\textrm{\scriptsize 29}$,
\AtlasOrcid[0000-0002-9987-2292]{A.~Aboulhorma}$^\textrm{\scriptsize 35e}$,
\AtlasOrcid[0000-0001-5329-6640]{H.~Abramowicz}$^\textrm{\scriptsize 151}$,
\AtlasOrcid[0000-0002-1599-2896]{H.~Abreu}$^\textrm{\scriptsize 150}$,
\AtlasOrcid[0000-0003-0403-3697]{Y.~Abulaiti}$^\textrm{\scriptsize 117}$,
\AtlasOrcid[0000-0002-8588-9157]{B.S.~Acharya}$^\textrm{\scriptsize 69a,69b,q}$
% ...
\bigskip
\\

% ...
$^{29}$Physics Department, Brookhaven National Laboratory, Upton NY; United States of America.\\
% ...
$^{35}$$^{(a)}$Facult\'e des Sciences Ain Chock, R\'eseau Universitaire de Physique des Hautes Energies - Universit\'e Hassan II, Casablanca;$^{(b)}$Facult\'{e} des Sciences, Universit\'{e} Ibn-Tofail, K\'{e}nitra;$^{(c)}$Facult\'e des Sciences Semlalia, Universit\'e Cadi Ayyad, LPHEA-Marrakech;$^{(d)}$LPMR, Facult\'e des Sciences, Universit\'e Mohamed Premier, Oujda;$^{(e)}$Facult\'e des sciences, Universit\'e Mohammed V, Rabat;$^{(f)}$Institute of Applied Physics, Mohammed VI Polytechnic University, Ben Guerir; Morocco.\\
% ...
$^{49}$Fakult\"{a}t Physik , Technische Universit{\"a}t Dortmund, Dortmund; Germany.\\
% ...
$^{55}$II. Physikalisches Institut, Georg-August-Universit\"{a}t G\"ottingen, G\"ottingen; Germany.\\
% ...
$^{69}$$^{(a)}$INFN Gruppo Collegato di Udine, Sezione di Trieste, Udine;$^{(b)}$ICTP, Trieste;$^{(c)}$Dipartimento Politecnico di Ingegneria e Architettura, Universit\`a di Udine, Udine; Italy.\\
% ...
$^{102}$CPPM, Aix-Marseille Universit\'e, CNRS/IN2P3, Marseille; France.\\
% ...
$^{117}$Department of Physics, New York University, New York NY; United States of America.\\
% ...
$^{120}$Homer L. Dodge Department of Physics and Astronomy, University of Oklahoma, Norman OK; United States of America.\\
% ...
$^{150}$Department of Physics, Technion, Israel Institute of Technology, Haifa; Israel.\\
$^{151}$Raymond and Beverly Sackler School of Physics and Astronomy, Tel Aviv University, Tel Aviv; Israel.\\
% ...
$^{a}$ Also Affiliated with an institute covered by a cooperation agreement with CERN.\\
$^{b}$ Also at An-Najah National University, Nablus; Palestine.\\
$^{c}$ Also at APC, Universit\'e Paris Cit\'e, CNRS/IN2P3, Paris; France.\\
$^{d}$ Also at Borough of Manhattan Community College, City University of New York, New York NY; United States of America.\\
$^{e}$ Also at Center for High Energy Physics, Peking University; China.\\
$^{f}$ Also at Center for Interdisciplinary Research and Innovation (CIRI-AUTH), Thessaloniki; Greece.\\
% ...
$^{q}$ Also at Department of Physics, King's College London, London; United Kingdom.\\
% ...

\end{flushleft}


\end{document}